\input epsf
\documentstyle{l-aa}

\newcommand{\beq}{\begin{equation}}
\newcommand{\eeq}{\end{equation}}
\newcommand{\eeql}[1]{\label{eq:#1}\end{equation}}

\newcounter{compteur}

\def\mv{M_V}
\def\mi{M_I}
\def\814{I_{814}}
\def\606{R_{606}}
\def\555{V_{555}}
\def\mh{[M/H]}
\def\feh{[Fe/H]}
\def\oh{[O/H]}
\def\msol{M_\odot}
\def\mso{M_\odot}

\def\te{T_{\rm eff}}
\def\simgr{\,\hbox{\hbox{$ > $}\kern -0.8em \lower 1.0ex\hbox{$\sim$}}\,}
\def\simle{\,\hbox{\hbox{$ < $}\kern -0.8em \lower 1.0ex\hbox{$\sim$}}\,}
\def\wig#1{\mathrel{\hbox{\hbox to 0pt{%
          \lower.5ex\hbox{$\sim$}\hss}\raise.4ex\hbox{$#1$}}}}

\def\aj{AJ}                  
\def\araa{ARA\&A}             
\def\apj{ApJ}                 
\def\apjl{ApJ}                
\def\apjs{ApJS}               
\def\aap{A\&A}                
\def\mnras{MNRAS}             
\def\pasp{PASP}               

\begin{document}
\thesaurus{}
\title{
Evolutionary models for metal-poor low-mass stars. Lower
main sequence of globular clusters and halo field stars}

\author{{\sc I. Baraffe\inst{1}, G. Chabrier\inst{1}, F. Allard\inst{2} and
 P. H. Hauschildt\inst{3}}}

\institute{C.R.A.L. (UMR 5574 CNRS),
Ecole Normale Sup\'erieure, 69364 Lyon Cedex 07, France,
(ibaraffe, chabrier @ens-lyon.fr)
 \and Dept. of Physics, Wichita State University,
Wichita, KS 67260-0032,
(allard@eureka.physics.twsu.edu) \and Dept. of Physics and Astronomy, University of Georgia
Athens, GA 30602-2451,
(yeti@hal.physast.uga.edu)}

\date{Received date ; accepted date}

\maketitle

\markboth{I. Baraffe et al.:Evolutionary models for metal-poor low-mass stars } {}

\begin{abstract}

\bigskip
We have performed evolutionary calculations of very-low-mass stars
from 0.08 to 0.8 $\msol$ for different metallicites from [M/H]= -2.0
to -1.0 and we have tabulated the mechanical, thermal and photometric
characteristics of these models. The calculations include the most
recent interior physics and improved non-grey
atmosphere models.  The models reproduce the entire main
sequences of the globular clusters observed with the Hubble Space
Telescope over the afore-mentioned range of metallicity. Comparisons
are made in the WFPC2 {\it Flight} system including the F555, F606 and
F814 filters, and in the standard Johnson-Cousins system.  We examine
the effects of different physical parameters, mixing-length,
$\alpha$-enriched elements, helium fraction, as well as the accuracy
of the photometric transformations of the HST data into standard
systems. We derive mass-effective temperature and mass-magnitude
relationships and we compare the results with the ones obtained with
different grey-like approximations. These latter are shown to yield
inaccurate relations, in particular near the hydrogen-burning
limit. We derive new hydrogen-burning minimum masses, and the corresponding
absolute magnitudes, for the different
metallicities.

We predict color-magnitude diagrams in the infrared NICMOS filters,
to be used for the next generation of the HST observations, providing
mass-magnitudes relationships in these colors down to the brown-dwarf
limit. We show that the expected signature of the stellar to
substellar transition in color-magnitude diagrams is a severe
blueshift in the infrared colors, due to the increasing
collision-induced absorption of molecular hydrogen with increasing
density and decreasing temperature.

At last, we apply these calculations to the observed halo field stars,
which yields a precise determination of their metallicity, and thus of
their galactic origin. We find no evidence for significant
differences between the halo field stars and the globular cluster
sequences.

\bigskip

Subject headings : stars: Low-mass, brown dwarfs --- stars: evolution
--- stars: globular clusters
\end{abstract}

\section{Introduction}

Over the past decade considerable effort, both observational and
theoretical, has been directed towards a more accurate determination
of the stellar lower main sequence, down to the edge of the
sub-stellar domain. Such a determination bears important consequences
for our understanding of a wide variety of astrophysical problems,
from star formation and stellar structure to galactic formation and
evolution.  Although the lower main sequence of the disk is relatively
well determined since the survey of Monet et al. (1992), the situation
is less well defined for the Galactic halo, primarily because of the
greater difficulties involved in identifying the
halo stars.
The task of detecting halo low-mass stars (LMS) and measuring their
magnitude is of formidable difficulty with ground-based
telescopes. Globular clusters (GCs) have always presented a particular
interest for the study of the stellar halo since we can more easily
determine their main sequence than for field halo stars.  Most of the
observations of GCs focussed on the {\it upper} main sequence,
i.e. the turn-off point, and the red giant branch, for a comparison of
this region with theoretical isochrones yields a determination of the
age of the clusters, and thus a lower-bound for the age of the Galatic
halo.  Thanks to the tremendous progress in deep photometry realized
recently with the Hubble Space Telescope (HST), which reaches
unprecedented magnitude and spatial resolution, the lower main
sequence of GCs is now observed nearly down to the hydrogen burning
limit.  Thanks to the high angular resolution achievable with the HST,
accurate photometry is feasible to levels about 4
magnitudes fainter than with ground-based observations, allowing
photometry of very faint stars.  Several HST observations
of globular clusters are now available, spanning a large
metallicity-range from solar value to substantially metal-depleted
abundances, as will be presented in the next section. The lower main
sequence of these clusters is well defined and offers a unique
possibility to probe low-mass star evolutionary models for various
metallicities down to the hydrogen-burning limit.

In spite of considerable progress in stellar theory - internal
structure, model atmospheres and evolution - all the LMS models so far
failed to reproduce accurately the observed color-magnitude diagrams
(CMD) of disk or halo stars below $\sim 4000$ K, i.e. $\sim
0.4-0.6\,\msol$, depending on the metallicity. All the models
predicted too hot an effective temperature for a given luminosity,
i.e. were too blue compared to the observations by at least one
magnitude (see e.g. Monet, 1992).
Such a disagreement stemed essentially from shortcomings both in the
physics of the interior, i.e. equation of state (EOS) and thus
mass-radius relationship and adiabatic gradient, and in the atmosphere, since all models were
based on grey atmospheres and a diffusion approximation. Important
progress has been made recently in this field with the derivation of
an appropriate EOS for low-mass stars and brown dwarfs (Saumon, Chabrier
\& Van Horn 1995; see Chabrier and Baraffe 1997), non-grey model atmospheres for M-dwarfs (Allard and
Hauschildt 1995, 1997; Brett 1995) and evolutionary models based on a
consistent treatment between the interior and the atmosphere profile
(Baraffe, Chabrier, Allard \& Hauschildt 1995; Chabrier, Baraffe \&
Plez 1996; Chabrier \& Baraffe, 1997). These models now reach quantitative agreement with the
observations, as shown in the afore-mentioned papers and presented in
\S 4.

Globular clusters offer the great advantage of all stars having
the same metallicity, determined relatively accurately from
bright star spectroscopic measurements.  Moreover they
are old enough for all the stars to have reached thermal
equilibrium, so that age effects do not affect the luminosity of the
objects near the bottom of the MS. For these reasons, the
mass-luminosity relationship is well determined along the entire MS of
GCs, from the turn-off down to the brown dwarf limit, with no
dispersion due either to age or metallicity.  From the theoretical
viewpoint, these properties restrain appreciably the degrees of
freedom in the parameter space, so that globular clusters provide a
very stringent test to probe the validity of low-mass star
evolutionary models and of the related mass-luminosity relationships
for various metallicities. Agreement between these models and
observations is a {\it necessary condition} (but not sufficient !) to
assess the validity of these mass-luminosity relationships, a
cornerstone to derive {\it reliable} mass-functions.

In this paper, we present new evolutionary models for metal-depleted
($[M/H]\le -1$) low-mass stars ($m\le 0.8\,\msol$), based on the most
recent non-grey model atmospheres.  We compare the results with the
observed CMDs of three globular clusters, namely $NGC6397$, $M15$
and $\omega Cen$, for which HST observations are available. The paper
is organized as follows : the observations are summarized in \S 2
whereas the theory is outlined in \S 3, where comparison is made with
other recent LMS models. Comparison between theory and observation is
presented in \S 4. Section 5 is devoted to discussion and conclusion.

The extension of the present calculations to more metal-rich clusters ([Fe/H]
$\ge$ -0.5)
and field stars, which require more extensive calculations, will be
presented in a forthcoming paper (Allard et al. 1997b), as well as the
derivation of the mass function for globular star clusters and halo
field stars, from the observed luminosity functions (Chabrier \& M\'era,
1997).

\section{Observations}

The decreasing luminosity of stars with mass ($L\sim m^{2.5-3.5}$
for low-mass stars) renders the observation of the lower main sequence
almost impossible from ground-based telescopes. Despite these
difficulties, a number of clusters have been investigated with large
telescopes by several groups worldwide to determine the bottom of the
main sequence.  To our knowledge, the most extensive study was
conducted by Richer et al. (1991) who determined accurately the MS of
six GCs up to a magnitude $V \sim 24; \mv\sim 10$, i.e. a mass $m \sim
0.4\,\msol$. Although intrinsically interesting, these observations do
not provide any information about the very bottom of the MS, and the
shape of the luminosity function near the hydrogen-burning
limit. Moreover, for metal-depleted abundances, the afore-mentioned
limit magnitude and mass correspond to a temperature $\te \sim 4000$ K
(see \S 4). Above this temperature, the physics of the stellar
interiors and atmospheres is relatively well mastered, so that these
observations put little constraint on the models, and the related
mass-luminosity relationship.

A recent breakthrough in the field has been accomplished with the deep
photometry of several GCs obtained with the refurbished Space
Telescope. Several observations are now available for $NGC6397$
(Paresce, DeMarchi \& Romaniello, 1995; Cool, Piotto \& King, 1996),
$NGC7078$ ($\equiv M15$) (DeMarchi \& Paresce 1995; Piotto, Cool \&
King, 1996)
 and
$\omega \,Cen$ (Elson, Gilmore, Santiago \& Casertano, S., 1995).
Observations for these clusters reach $V\sim 27$, $I\sim 24$,
almost the very bottom of the main sequence, and thus provide a unique
challenge to probe the validity of the LMS models down to the brown
dwarf regime.

{$\bullet $} {\it Metallicity}. All the afore-mentioned clusters are
substantialy metal-depleted. Before going any further, it is essential
to define what we call "metallicity" in the present context, for
different definitions are used in the literature. What is observed in
globular clusters is the iron to hydrogen ratio $\feh$. The continuous
production of oxygen in type II supernovae during the evolution of the
Galaxy has led to the well observed enhancement of the oxygen to
iron $[O/Fe]$ abundance ratio in old metal-poor stars with respect to
the young disk population. Since our basic models assume a solar-mix
composition, i.e. $[Fe/O]=0$, the afore-mentioned oxygen-enhancement
must be taken into account to make consistent comparison between
theory and observation.
We use the prescription derived by Ryan and Norris (1991)
for halo subdwarfs, i.e.

$$\mh \, \approx [O/H] = [O/Fe]\,+\, \feh \eqno(1)$$

with

$$[O/Fe]\,=\,+0.35\,\,\,\,\,\,\,\,\,\,\,{\rm for}\,\,\, \feh\le -1$$
$$[O/Fe]\,=\,-0.35\times \feh \,\,\,\,{\rm for}\,\, -1\le\feh\le 0$$

Thus a cluster with an observed $\feh=-1.3$, for example, corresponds
to a model with a metallicity $[Z]=\log(Z/Z_\odot)=\mh\sim -1.0$,
certainly not $\mh=-1.3$. This latter choice does not take into
account the enhancement of the $\alpha-$elements\footnote{The
$\alpha$-elements include $O$,$Ne$,$Na$,$Mg$,$Al$,$Si$,$P$,$S$,$Cl$,$Ar$,$Ca$
and $Ti$}
and yields inconsistent comparisons, especially at the bottom of the
MS where the stellar optical spectra and colors are shaped by
H$^-$, MgH, CaH and TiO opacities.  The Ryan \& Norris
correction is based on the spectroscopically-determined abundance of
370 kinematically selected halo stars in the solar neighborhood.
Although this procedure is not as fully consistent as calculations
conducted with the exact mixture, it certainly represents a fairly
reasonable correction for the metal-poor star oxygen enrichment. The
accuracy of this prescription will be demonstrated in \S3.4, where
calculations with the appropriate $\alpha$-enhanced mixture are presented.

{$\bullet $} {\it Photometric conversion}.  The afore mentioned
clusters have been observed with the {\it Wide Field and Planetary
Camera-2 (WFPC2)} of the HST, using either of the $F814W\sim I$,
$F555W\sim V$ or $F606W$ filters, where $I,V$ refer to the standard
(Johnson-Cousins) filters. Thanks to the courtesy of A. Cool, I. King,
G. DeMarchi, F. Paresce, G. Gilmore and R. Elson, we have been able to
use the data in these filters, in the so-called WFPC2 {\it Flight}
system for $NGC6397$, $M15$ and $\omega \,Cen$. For these clusters,
comparison between theory and observations is made directly in the
{\it Flight} system, thus avoiding any possible uncertainties
in the {\it synthetic} flight-to-ground photometric transformations
of Holtzman et al. (1995).
The model magnitudes were calculated using the filter transmissions
curves and the {\it observed}\footnote{Since {\it observed} zero-points
were not available for the F606W filter, we used the {\it synthetic}
value listed in table 9 of Holtzman et al. (1995) for this filter.
An inspection of this table confirms that {\it synthetic} and
{\it observed} zero points agree almost exactly for the standard
WFPC2 filters.} zero points prescribed by Holtzman et al. (1995).

The conversion of apparent $m$ (observation) to absolute $M$ (theory)
magnitudes requires the knowledge of the distance modulus ($m-M$),
corrected for the interstellar extinction in each filter. In
order to minimize the bias in the comparison between theory and
observation, we have used the analytical relationships of Cardelli et
al. (1989) to calculate the extinctions from the M-dwarf synthetic
spectra of Allard \& Hauschildt (1997) over the whole frequency-range,
from the reddening value $E(B-V)$ quoted by the observers,
and we have compared the observed data with the theoretical models
{\it corrected for reddening}
in the WFPC2 filters, when available. The
extinction in each filter was found to depend very weakly on the
spectral type ($\wig < 0.05$ mag), and thus to be fairly constant over
the whole considered temperature range.  We believe these
determinations, based on accurate synthetic spectra, to yield the most
accurate extinction and reddening corrections and the most consistent
comparison between theory and observation. These values are given in
Table I, for each cluster, and compared to the values quoted by the
observers.  Table I also summarizes the characteristics adopted for
the three clusters of interest.  Note that some undetermination remains
in the distance modulus of $NGC6397$, yielding a difference in the
magnitude of $\sim 0.2$ mag, as will be shown in \S 4. This stresses
the need for a more accurate determination of this parameter.

\section{Theory}

We have derived recently evolutionary models aimed at describing the
mechanical and thermal properties of LMS. These models are based on
the most recent physics characteristic of low-mass star interiors,
equation of state (Saumon, Chabrier \& VanHorn 1995), enhancement
factors of the nuclear rates (Chabrier 1997) and updated opacities
(Iglesias \& Rogers 1996; OPAL), the last generation of non-grey
atmophere models (Allard \& Hauschildt 1997, AH97) and accurate
boundary conditions between the interior and the atmosphere
profiles. This latter condition is crucial for an accurate description
of LMS evolution, for which any diffusion approximation or grey
treatment yield inaccurate results (Baraffe, Chabrier, Allard \&
Hauschildt 1995; Chabrier, Baraffe and Plez 1996; Chabrier \& Baraffe, 1997).
A complete
description of the physics involved in these stellar models is
presented in Chabrier and Baraffe (1997) and we refer the reader to
this paper for detailed information.

Below $\sim 0.4\, \msol$, the stellar interior becomes fully
convective : the evolution of LMS below this limit is rather
insensitive to the mixing length parameter $\alpha = l/H_p$, and thus
the models are not subject to {\it any} adjustable parameter. From
this point of view, VLMS represent a formidable challenge for stellar
evolution theory.  Comparison with the observations is straightforward
and reflects directly the accuracy of the physics and the treatment of
the (self-consistent) boundary conditions involved in the
calculations. Any VLMS model including an adjustable parameter or a
grey atmosphere approximation in order to match the observations would
reflect shortcomings in the theory and thus would lead to unreliable
results. From this point of view, it is important to stress that,
although agreement with the observation is a {\it necessary} condition
to assess the validity of a stellar model, it is not a {\it
sufficient} condition. This latter requires the assessment
of the accuracy of the input physics and requires {\it
parameter-free, self-consistent} calculations.

The first generation of the present models (Baraffe et al. 1995) were
based on the "Base" grid of model atmospheres of Allard and Hauschildt
(1995; AH95).  These models improved significantly the comparison with
the observed Pop I and Pop II M-dwarfs sequences (Monet et al. 1992),
down to the bottom of the main-sequence, w.r.t. previous models
(Baraffe et al. 1995).  The AH95 models have been improved recently
by including (i) a pressure broadening treatment of the lines,
(ii) more complete molecular line lists, and (iii) by extending the
{\it Opacity-Sampling} technique to the treatment of the molecular
line absorption coefficients.  This yields the so-called "NextGen"
models (
Allard and Hauschildt 1997).  The "NextGen" synthetic spectra and
colors have been compared with Population-I M-dwarf observations by
Jones et al.  (1995, 1996), Schweizer et al. (1996), Leggett et
al. (1996), Viti et al. (1996), and have been used for the analysis of
the brown dwarf Gl229B by Allard et al. (1996).  The present work is
the first application of the "NextGen" model atmospheres to
observations and evolutionary calculations of metal-poor populations.

A first (preliminary) set of the present improved  stellar models
has been shown to reproduce accurately the observed mass-magnitude
(Henry \& Mc Carthy 1993) and mass-spectral type (Kirkpatrick \& Mc
Carthy 1994) relationships for solar-metallicity LMS down to the bottom of the main
sequence (Chabrier, Baraffe and Plez 1996; Baraffe \& Chabrier 1996).
The aim of the present work is to extend the comparison between theory
and observation to metal-depleted abundances, characteristic of old
disk and halo populations.

For this purpose, we have calculated evolutionary models for masses
$m\le 0.8\,\msol$ down to the hydrogen-burning limit, based on the
"NextGen" model atmospheres, for different low metal abundances
characteristic of old globular clusters, namely $\mh=-2,\,-1.5,\,
-1.3$ and $-1.0$.  This low-metal grid represents a first step towards
a complete generation of models covering solar metallicities.  A major
source of absorption in VLMS arises from the presence of $TiO$ and
$H_2O$ molecular bands in the optical and the infrared, respectively.
Although tremendous progress has been accomplished over the past few
years in this type of calculations, some uncertainty remains in
the absorption coefficients of these molecules, which affect not only
the spectrum, i.e. the colors, but also the profile of the atmosphere,
and thus the evolution of late-type M dwarfs (see Allard et al.
1997a for a review of VLM stellar atmospheres modeling).
Furthermore, the onset of grain formation, not included in the present
models, may also affect the spectroscopic and structural properties of
metal-rich VLMS below $\te \approx 2600$~K (Tsuji et al. 1996a,b;
Allard et al. 1996).  We expect these shortcomings in the models to be
of decreasing importance with decreasing metallicity.  Strong double-metal
bands ($TiO$, $VO$), for example, are noticeably weaker in the
spectra of metal-poor subdwarf (see e.g. Leggett 1992; Dahn et al. 1995),
whereas they would dominate the spectrum of a solar metallicity object at
this temperature.  For this reason it is important to first examine the
accuracy of the models for low-metal abundances and its evolution with
increasing metallicity.  The observed MS of the three afore-mentioned
HST GCs represent a unique possibility to conduct such a project.

In some cases, comparison is made with the first generation of models
(Baraffe et al. 1995), based on the "Base" model atmospheres. This
will illustrate the effect of the most recent progress accomplished in
the treatment of LMS atmospheres.  All the basic models assume a
mixing length parameter, both in the interior and in the atmosphere,
$\alpha=l/H_P= 1.0$.  Since the value of the mixing length parameter
is likely to depend to some extent on the metallicity, there is no
reason for the value of $\alpha$ in globular clusters to be the same
as for the Sun. Although the choice of $\alpha$ is inconsequential for
fully convective stars, it will start bearing consequences on the
models as soon as a radiative core starts to develop in the interior.
For metallicities $\mh \le -1$, this occurs for masses $m\ge 0.4$
$\msol$ (see Chabrier \& Baraffe 1997 for details).  In order to
examine the effect of the mixing length parameter for masses above
this limit, we have carried out some calculations with $\alpha =2$,
both in the interior and in the atmosphere (see \S 3.2 below).  All
the models assume a solar-mix abundance (Grevesse \& Noels 1993).  As
discussed in the previous section, oxygen-enhancement in metal-poor
stars is taken into account by using the eqn.(1) as the correspondence
law between solar-mix and $\alpha$-enhanced abundances.  In order to
test the accuracy of this procedure, a limited set of calculations
with the exact $\alpha$-enriched mixture have also been performed (\S
3.4).

The adopted helium fraction is $Y=0.25$ for all metallicities.  The
effect of helium abundance will also be examined below.

All models have been followed from the initial deuterium burning phase
to a maximum age of 15 10$^9$ yr.

Before comparing the results with observations, we first examine some
intrinsic properties of the models. Comparison is also made with other
recent LMS models for metal-poor stars, namely D'Antona \& Mazzitelli
(1996; DM96) and the Teramo group of Dr. Castellani (Alexander et
al. 1997). Although the EOS used by the Teramo group is the same as
in the present calculations (SCVH), both groups use grey model
atmospheres based on a $T(\tau)$ relationship, and match the interior
and the atmosphere structures either at an optical depth $\tau=2/3$
(DM96) or at the onset of convection (Teramo group).

\subsection{Mass-effective temperature relationship}

Figure 1 shows the mass-effective temperature relationship for two
metallicities, $\mh=-1.5$ ($Z=6\times 10^{-4}$) (solid line) and $\mh=-1.0$
($Z=2\times 10^{-3}$)(dash-dot line). We note the two well established changes in the
slope at $\te \approx$ 4500 K and 3800 K, respectively, for these metallicities.
The first one corresponds to the onset of molecular
formation in the atmosphere, and the related changes in the opacity
(see e.g. Copeland, Jensen \& Jorgensen 1970; Kroupa, Tout \& Gilmore
1990). The second reflects the overwhelming importance of electron
degeneracy in the stellar interior near the hydrogen burning limit
(DM96; Chabrier \& Baraffe 1997).  As expected, the enhanced pressures
of lower metallicity mixtures yield larger effective temperatures
for a given mass (see Chabrier \& Baraffe, 1997).

\begin{figure}
\epsfxsize=88mm
\epsfysize=90mm
\epsfbox{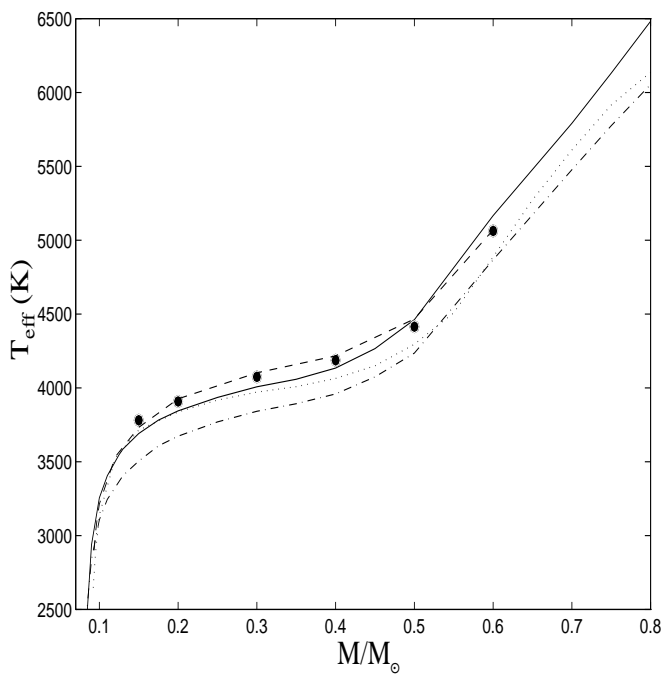} 
\caption[ ]{ Mass - effective temperature relation for $\mh$=-1.5
(solid line) and $\mh$=-1.0 (dash-dot line). The full circles denote our
calculations done with the Krisha-Swamy grey-like treatment (see text) for
$\mh$=-1.5 while the dashed line show the results of the Teramo group
(Alexander et al., 1997) for the same metallicity. The dotted line shows the results
of D'Antona \& Mazzitelli (1996; DM96) for $\mh$=-1.0.}
\end{figure}

Also shown on the figure are comparisons with DM96 and the Teramo
group, for comparable metallicities.  The DM96 models (dotted line)
overestimate substantially the effective temperature, by up to $\sim 200
$ K below 0.5 $\msol$.  This will yield substantially blue-shifted
sequences in the CMDs.  This is a direct consequence of using a
$T(\tau)$ prescription (and a different EOS, and adiabatic gradient),
which implies a grey atmosphere and radiative equilibrium, two conditions
which become invalid under low-mass star characteristic conditions, as
demonstrated in previous work (Chabrier, Baraffe \& Plez 1996;
Chabrier \& Baraffe 1997). The Krishna-Swamy (1966,KS) formula used by the
Teramo group is known to slightly improve this shortcoming.
As  shown
in the figure, we reproduce the Teramo results (dashed line)
when using the KS $T(\tau)$ relationship (filled circles).
Though smaller, the departure of the mass-$\te$ relationship in that case
is still significant, in particular near the brown dwarf domain, where it
becomes too steep and predicts too large hydrogen burning minimum masses
(see Chabrier \& Baraffe, 1997).
A detailed examination of these different prescriptions is given in
Chabrier \& Baraffe (1997).  These comparisons show that,
even though some improvement can be reached w.r.t. the basic
Eddington approximation, {\it any} atmosphere-interior boundary
treatment based on a prescribed $T(\tau)$ relation
yields incorrect, substantially overestimated, effective temperatures
for $\te\wig < 4500$ K, i.e.  $m\wig <0.5\,\msol$. Conversely, they
underestimate the mass for a given temperature (luminosity) below this
limit, i.e. at the bottom of the MS. This
illustrates the unreliability of any grey-like
treatment to describe accurately the non-grey effects and the presence
of convection in optically thin layers of VLM stars.
This bears important
consequences for the cooling history and thus the evolution in
general, and for the mass-luminosity relationship and thus the mass
calibration in particular.

\subsection{Effect of the mixing length}

In this section, we examine the effect of the mixing length parameter
$\alpha=l/H_p$ on the stellar models, for a given metallicity.  We
must distinguish the effect of this parameter in the stellar {\it
interior} and in the {\it atmosphere}. Both effects reflect the
importance of convective transport in two completely different regions
of the star, namely the envelope and the photosphere, and thus do not
necessarily bear the same consequences on the models.

We first examine the effect of the mixing length parameter in the {\it
atmosphere}, by conducting evolutionary calculations with non-grey
model atmospheres for [M/H]=-1.5, calculated with $\alpha=1$ and 2,
while the value is kept unchanged in the interior.  We have selected a
range of effective temperatures and gravities ($T_{eff}$ = 4000 - 5800
K, log g = 4.5 - 5) corresponding to masses $m = 0.3 - 0.7$ $M_\odot$.
The atmosphere profiles corresponding to both situations are shown in
Fig. 2, where the onset of convection and the location of the optical
depth $\tau$ = 1 are indicated. As seen on the figure, the atmosphere
profile is rather insensitive to a variation of $\alpha$ in the {\it
optically-thin} region, as expected from the rather shallow convection
zone in the atmosphere at this metallicity and effective temperature.
The main consequence is that the spectrum, and thus the colors and
magnitudes at a given $T_{eff}$, remain almost unaffected (less than
0.04 mag in the afore-mentioned range).  Below $\sim 3000$ K, the
effect of the mixing length in the atmosphere is inconsequential (see
also Brett 1995).  The evolutionary models calculated with both sets
of model atmospheres in the selected range of effective temperatures
differ by less than 50 K in $\te$. The effect of the mixing length in
the atmosphere thus bears no consequence on the evolution.
\begin{figure}
\epsfxsize=88mm
\epsfysize=88mm
\epsfbox{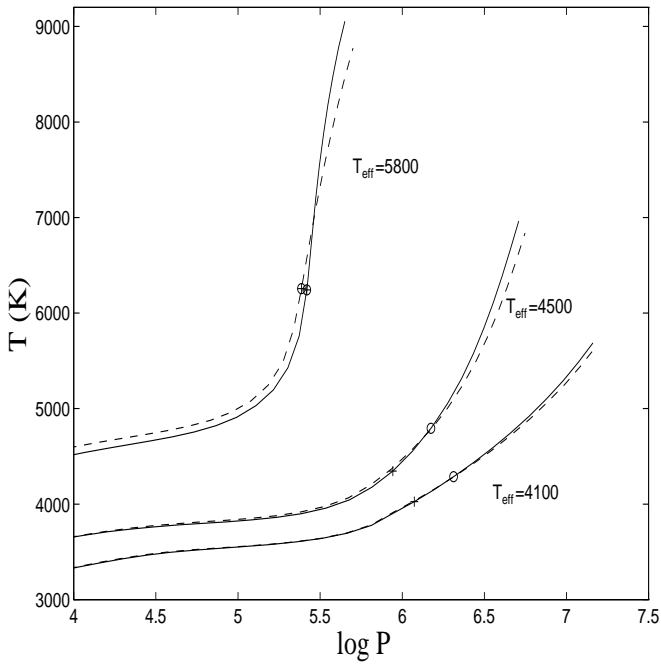} 
\caption[ ]{ Atmosphere profiles for $\mh$=-1.5 and a mixing length
 $l=H_P$ (solid line) and $l=2\,H_P$ (dash line), respectively, for selected
effective temperatures. The open circles indicate the location of $\tau = 1$
and the crosses show the onset of convection.}
\end{figure}

We now examine the effect of convection in the stellar {\it interior},
by conducting calculations with $\alpha=1$ and 2 in the interior,
keeping the fiducial value $\alpha=1$ in the atmosphere. This is
illustrated in Figure 3a by comparing the solid line (and empty
circles), which corresponds to $\alpha=1$ in the interior, with the
dashed line (triangles), which corresponds to $\alpha=2$, for $t=10$
Gyr.  As expected the effect is null or negligible for
convection-dominated interiors, i.e. $m\wig < 0.6\,\msol$. Above this
limit, a larger mixing parameter, i.e. a more efficient convection,
yields slightly hotter (bluer) models (D'Antona \& Mazzitelli 1994;
Chabrier \& Baraffe, 1997),
about 200-300 K for $\te =5800-6500$ K, i.e. $m=0.7-0.8\,\msol$, for
the metallicity of interest, $\mh=-1.5$.

In summary, the effect of the variation of the mixing length parameter
on the stellar models remains weak for the lower MS and is essentially
affected by the value in the stellar interior.  A variation of
$\alpha=1$ to 2 yields variations $\Delta \te/\te\sim 2-5 \%$ on the
effective temperature in the upper MS, for masses above $\sim
0.6\,\msol$, while the luminosity remains essentially unaffected.  We
will examine in section 4.1 whether comparison with the observed MS of
GCs can help calibrating this value.

\subsection{Effect of the age}

The effect of the age in a color-magnitude diagram is shown on Figure 3a, for
$\mh=-1.5$ for ages $t=10$ and 15 Gyr.  The time required to reach the
zero-age main sequence is much smaller than the age of GCs over the
entire stellar mass range so that {\it all}
hydrogen burning objects do lie on the MS and the bottom of the GC MS
is unaffected by age variation.  On the other hand, metal-poor stars
are significantly hotter and more luminous than their more metal-rich
counterparts.  Therefore, for a given mass, they burn more rapidly
hydrogen in their core, and thus evolve more quickly off the MS. The
more massive (i.e.  hotter) the star, the larger the effect. This
defines the turn-off point, i.e. the top of the MS. For fixed
metallicity, the turn-off point will thus be reached for lower masses,
i.e. fainter absolute magnitudes, as the age increases.  This is
illustrated in Figure 3a where we compare the 10 Gyr and 15 Gyr
isochrones calculated for different mixing length parameters, for the
same metallicity.  While the position in the HR diagram of masses
below $\sim 0.6\,\msol$ remains unchanged, larger masses become
substantially bluer and more luminous with age as they transform more
central hydrogen into helium. The resulting increase of molecular
weight yields further contraction and heating of the central
layers, and thus an increase of the radiative flux ($F_{rad}\propto
T^3$) and the luminosity.  Eventually they evolve off the MS for a
(turn-off)-mass $m_{TO}\sim 0.8 \,\msol$ for
$\mh=-1.5$ and $-2$ at t=10 Gyr and $m_{TO}\sim 0.75
\,\msol$ at 15 Gyr.  As shown on the figure, the upper MS is sensitive
to age and mixing length variations. The calibration of the mixing
length parameter on the turn-off point is then altered by age
uncertainties.  The {\it precise} determination of
the turn-off point thus requires more detailed calculations, which are out
the scope of the present study.
\begin{figure}
\epsfxsize=88mm
\epsfysize=88mm
\epsfbox{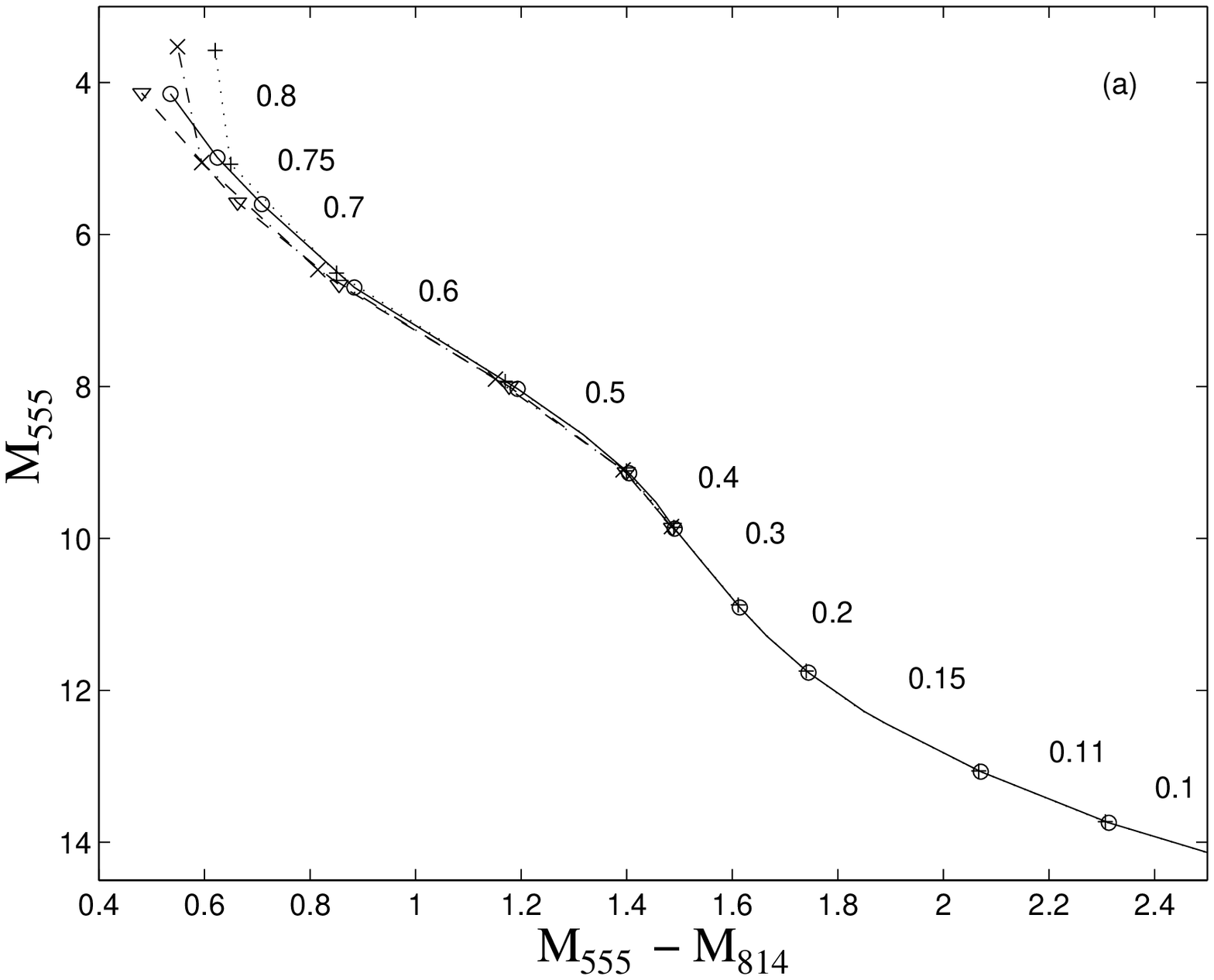}
\epsfxsize=88mm
\epsfysize=88mm
\epsfbox{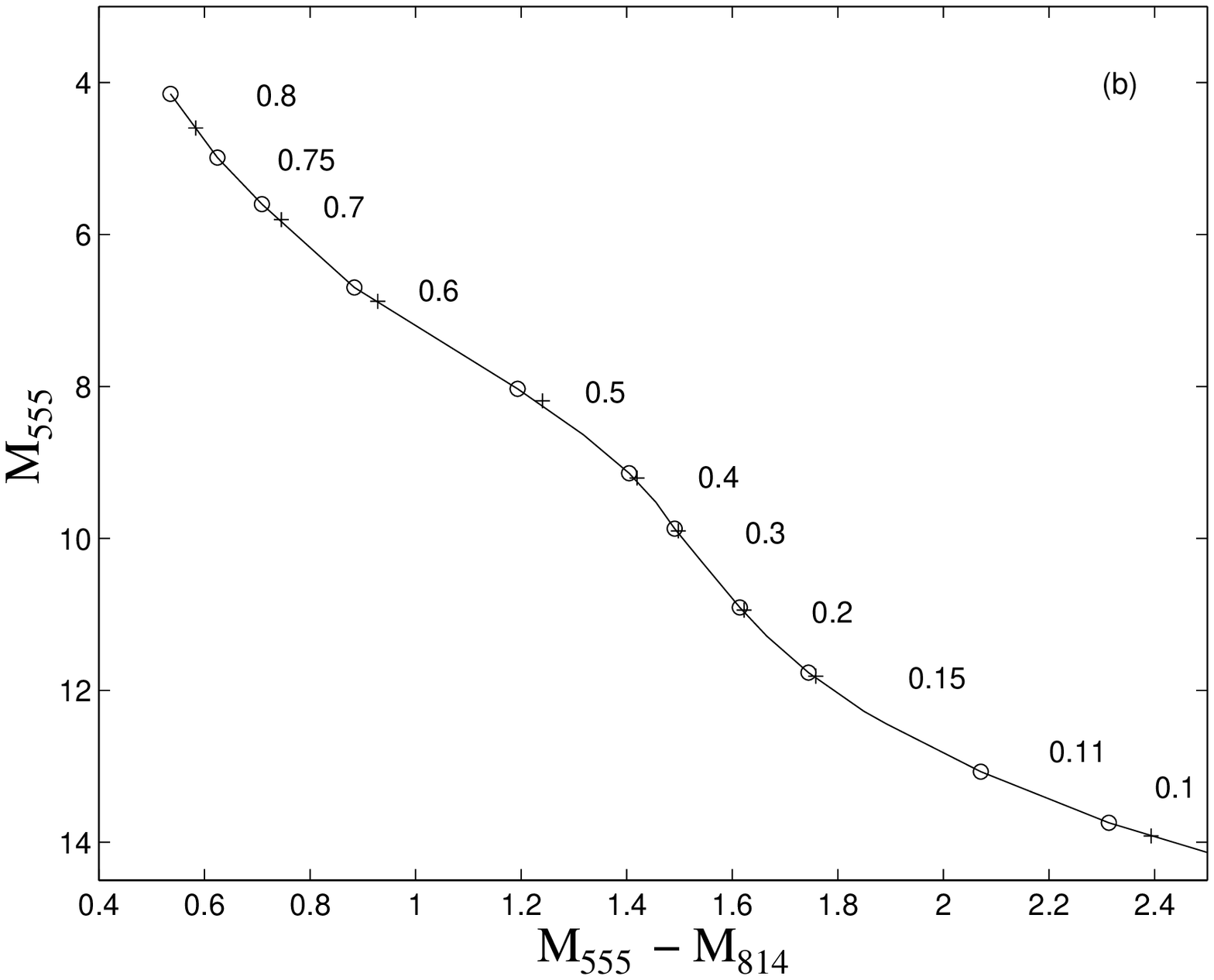}  
\caption[ ]{{\bf (a)} Effect of the age and of the  mixing length in the stellar interior
for [M/H]=-1.5.
 The
models display the $t=10$ Gyr isochrone when using a mixing length $l=H_P$
(solid line; empty circles) and $l=2\,H_P$ (dashed line; triangles).
The dotted line ($+$) shows the 15 Gyr isochrone calculated with $l=H_P$,
whereas the dash-dotted curve ($X$) corresponds to t=15 Gyrs and
$l=2 \,H_P$. The masses indicated on the figure correspond to the 10 Gyrs
isochrones (open circles and triangles). The upper mass corresponding to the 15 Gyrs isochrones is 0.75 $\mso$. {\bf (b)} Effect of the helium fraction $Y$, for $\mh=-1.5$,
with Y = 0.25 (solid curve, open circles) and Y = 0.23 (+).
 }
\end{figure}

\subsection{Effect of helium and $\alpha$-elements}

As mentioned above, the fiducial calculations have been conducted with
a helium abundance $Y=0.25$. Figure 3b compares these results (solid
line; circles), for $\mh=-1.5$, with calculations done with $Y=0.23$
($+$).  The corresponding variations of the temperature (color) and
the luminosity (magnitude) are negligible over the entire mass range,
except near the turn-off point.

In order to examine the accuracy of the Ryan \& Norris prescription
(eqn.(1)), we have conducted calculations using model
atmospheres computed with an $\alpha$-element abundance
enrichment of $[\alpha/Fe]=0.3$ for a $\feh=-1.3$ ($[\alpha/H]=-1.0$)
mixture.
Since
the spectra of VLMS near the bottom of the MS are governed by
$\alpha$-elements, we expect this mixture to yield equivalent results
to those obtained with the scaled-solar abundance mixture of [M/H]=
-1.0. This test will also determine whether neglecting a relative
under-abundance of $Fe$ and other non-$\alpha$ elements in the
scaled-solar abundances does affect or not the shape of the MS. We have
computed model atmospheres with the afore-mentioned modified
abundances for several temperatures and gravities in the range
$\te=3000-5000$ K and $\log g=4.5-5.0$. We find that the atmosphere
density and temperature profiles and the colors are
essentially undistinguishable from the ones calculated with
$[\alpha /Fe]=0.0$, $\mh(=\feh)=-1.0$, i.e. a scaled solar-mix,
throughout the entire atmosphere.

We have also examined the effect of $\alpha$-enriched abundances in
the {\it interior} opacities (OPAL), thanks to appropriate opacities
kindly provided by F. Rogers for $\mh$=-1.0 and $[\alpha/Fe]=0.3$.
Here too we find that the effect is negligible and does not
modify noticeably the isochrones.  This assesses the validity of the
Ryan \& Norris procedure to take into account the $\alpha$-element
enrichment for metal-poor stars, and demonstrates the consistency of
our prescription, i.e. comparing solar-mix models with $[\alpha/H]=\mh=-1.0$
to observations with $\feh=-1.3$.  Conversely, this demonstrates the
inconsistency of comparing observations and solar-mix models for
these stars with $\feh=\mh$, as done sometimes in the literature.



\bigskip

The theoretical characteristics of the present models, effective
temperature, luminosity, gravity, bolometric magnitude and magnitudes in $VRIJHK$ for several
metallicities and an age t = 10 Gyrs are given in Tables II - V.

\section{Results and discussion}

\subsection{Comparison with globular cluster main sequences}

Figures 4-6 show the main sequence CMDs of the three GCs mentioned in
\S2.  As mentioned previously,
the
CMDs are shown in the $M_{555}$ ($\sim V$), $M_{814}$ ($\sim I$),
and $M_{606}$ filters in the WFPC2 {\it Flight} system to avoid
possible errors due to uncertain photometric conversion into the
standard Johnson-Cousins system.
In all cases,
the main sequence is well defined down to $I \approx 25$, $V \approx
27$.  Below this limit, the observed MS dissolves into the field and
it becomes quite difficult to distinguish the cluster-MS from field
stars.

The theoretical MS for the appropriate metallicity, as described in \S
2, are superimposed to the observations in each figure.  For
comparison in the $Flight$ system, observed magnitudes have been dereddened with
the corrections derived from the synthetic spectra (cf. Table I).  The
first striking result is the excellent agreement between theory and
observation for the three clusters, spanning a fairly large metallicity
range from strongly metal-depleted abundances ($M15$) to a tenth of
solar metallicity ($\omega \,Cen$). In particular, the changes of the
slope in the observed MS are perfectly reproduced by the models, for
the {\it correct magnitude, color} and {\it metallicity}.  Since these changes stem from
the very physical properties of the stellar interior and atmosphere,
as discussed in the previous section, the present qualitative and {\it
quantitative} agreement assesses the accuracy of the
physical inputs in the present theory.  The masses corresponding to
these changes are indicated on the curves and their effective
temperatures are given in Tables II - V, for the various metallicities.

Let us now consider each cluster in turn in order of increasing
metallicity.

{$\bullet$} $M15$ (Fig. 4a,b) : This is the most metal-depleted HST
cluster presently observed, with $\feh\sim -2.26$ (Djorgovski 1993),
i.e.  $\mh\sim -2.0$ within the prescription adopted in the present
paper.  The MS of $M15$, as observed by DeMarchi \& Paresce (1995), is
shown in Figure 4a, with the distance modulus adopted by these
authors.  Comparison is made with our models with $\mh=-1.5$ (solid
line) and -2.0 (dashed line).  Although both models merge in the upper
MS, metallicity effect is clearly reflected in the lower MS where the
peak of the spectral distribution falls in the V bandpass, i.e. below
$m < 0.6\,\msol$, $M_{606} - M_{814} > 0.6$, and $\te < 5000$ K.  As
seen on the figure, the agreement with the observations over the whole
MS is excellent for $\mh=-2$ (the observationally-determined
metallicity) although the models appear to be slightly too red by
$\sim 0.05$ mag or overluminous by $\sim 0.2$ mag, in the
upper MS.  Even though this discrepancy is within the observational
error bars in $m_{606}$ and $m_{814}$, which range from $\pm 0.03$
mag for the upper MS to $\pm 0.17$ mag for the lower MS (DeMarchi,
private communication), we will examine the possibilities for this
disagreement to be real.  A slightly lower metallicity would leave the
upper MS almost unchanged, as seen from the two theoretical sequences
displayed on the figure.  Although the offset in the upper MS might be
compensated by a slightly larger mixing length (see \S 3.2), the fact
that it is quite constant along the entire sequence --- whereas the
mixing length only affects the upper MS (see \S 3.2) --- rather
suggests either an overestimated reddening $E(B-V)$ (by $\sim$ 0.05
mag) or an underestimated distance modulus.
Indeed, assuming $(m-M)_{814}=15.5$,
i.e. $+0.2$ mag w.r.t. the value quoted by DeMarchi \& Paresce, or a
reddening fainter by $\sim 0.05$ mag would bring theory and
observation into perfect agreement.  A 0.2 mag error on the distance
modulus corresponds to $\sim 1$ kpc, i.e.  $\sim$ 10\% of the
canonical value used presently (Djorgovski 1993).  Such an error
cannot be excluded for this remote cluster ($D\approx 10.5$ kpc).

 Note that we exclude an artificial offset in
the calibration
of the model magnitudes as source of discrepancy, since we use the
magnitude zero-points
kindly provided by
G. DeMarchi. However, a calibration problem of the data in the $F606W$ filter cannot
be excluded,
since  a disagreement appears as well in the comparison of NGC6397 observed
by the same group in the same filters. The disagreement however vanishes
for the same
cluster in $M_{505}$ and $M_{814}$  and for $wCen$ observed
by Elson et al. (1995) in $M_{606}$ and $M_{814}$ (see below). 
The latter agreement seems to reject an intrinsic problem of the atmosphere models in the spectral region covered by the $M_{606}$ passband.
\begin{figure*}
\epsfxsize=180mm
\epsfysize=160mm
\epsfbox{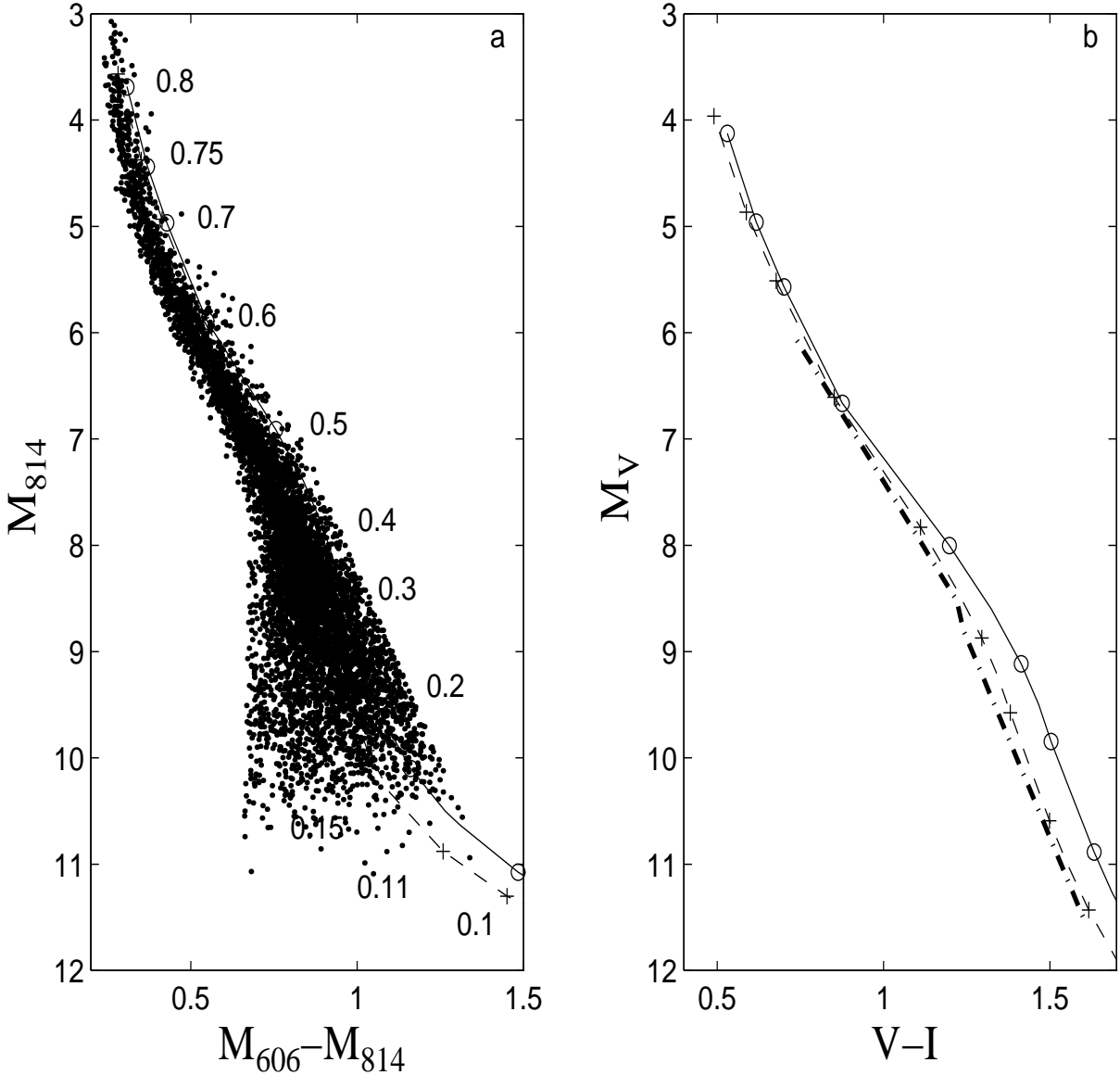} 
\caption[ ]{{\bf (a)}  CMD for $M15$.  The data in the $606$ and $814$ filters are from DeMarchi \& Paresce
(1995). The data are dereddened with the corrections calculated in the present study, as
given in Table I. Solid line (o): $\mh$=-1.5; dashed line (+): $\mh$=-2.0.
{\bf (b)} Same as in Fig. 4a in the Johnson-Cousins system with
the fit of Santiago et al. (1996) for the cluster (thick dash-dot).}
\end{figure*}

The data of DeMarchi \& Paresce have been transformed into the
standard $V-I$ Johnson-Cousins system by Santiago et al. (1996)
who used the dereddening correction quoted by the former
authors (see Table I) and the {\it synthetic} transformations of
Holtzman et al. (1995).  Figure 4b compares the $M15$ MS fit derived by
Santiago et al. to the present models in the same Johnson-Cousins
color system, where the filter transmissions of Bessell (1990) have
been applied to the models.
The agreement is similar to that obtained in the previous
{\it Flight} system, well within the error bars of the
photometric $m_{606}$-to-$V_J$ transformation, with the same slight offset
in color.
This comparison
assesses the validity of the photometric {\it synthetic}
transformations of Holtzman et al. (1995) in the present filters.

As shown in the figure and in Table II, the faintest observed stars on
the MS, $M_{814}\sim 10$ or $M_V \sim 11.5$, correspond to a mass
$m\sim 0.15\,\msol$, still well above the hydrogen burning limit. This
latter, $m=0.083\,\msol$ for $\mh=-2$, is expected to correspond to
$M_{814}\sim 14.6$, $M_{606} - M_{814} \sim 4.3$ or $M_V \sim 21$ and
$(V-I) \sim 6.5$ (Table II).  At the bright end, the limit of the
observations corresponds to the turn-off point $M_{814}\sim M_I\sim 3$
(DeMarchi \& Paresce 1995) which corresponds to $\sim 0.8\,\pm
0.05\,\msol$ depending on the age. Note that the Johnson-Cousins fit
given by Santiago et al. (1996) starts at $M_V \sim 6$, which
corresponds to a mass $m \sim 0.6 - 0.7 \msol$ and does not include
the upper MS.

{$\bullet$} $NGC6397$ (Fig. 5a,b,c) : This cluster has been observed
separately by Paresce et al. (1995) and Cool et al. (1996) in
different photometric filters, thus allowing comparison in the three
WFPC2 filters mentioned previously.  Figure 5a shows the comparison of
the observations and the models in $M_{814}$ vs $M_{606} - M_{814}$
for $\mh=-1.5$ and $-2$.
The best agreement is obtained for the
observed metallicity $\mh =-1.5$, although admitedly the theoretical sequence lies near the blue
edge of the low MS ($M_{814}\wig > 8$). We note however that the
error bars for this cluster range from $\pm 0.02$ for the
brightest stars to $\pm$ 0.20 mag for the lower MS (cf. Paresce et
al. 1995). The agreement displayed in Fig. 5a is therefore well
within the error bars.
Much better agreement is found in $M_{555}$ vs
$M_{555} - M_{814}$, as shown on Figure 5b, although an offset
now appears for the intermediate part of the sequence ($M_{555}\sim 8$),
$\sim 0.05$ mag in color and $\sim 0.3$ mag in magnitude.
As shown on
Figure 3a, this part of the MS is insensitive to the mixing length,
and the disagreement would thus not be solved by a larger mixing
length. In the same vein, models with $\alpha$-enriched abundances
would yield similar results, as discussed in \S 3.4. Models with a
substantially lower metallicity would fail reproducing the shape of
the MS both in $M_{555}$ (cf. Fig. 5b) and $M_{814}$ (see Fig. 5a).
\begin{figure*}
\epsfxsize=180mm
\epsfysize=100mm
\epsfbox{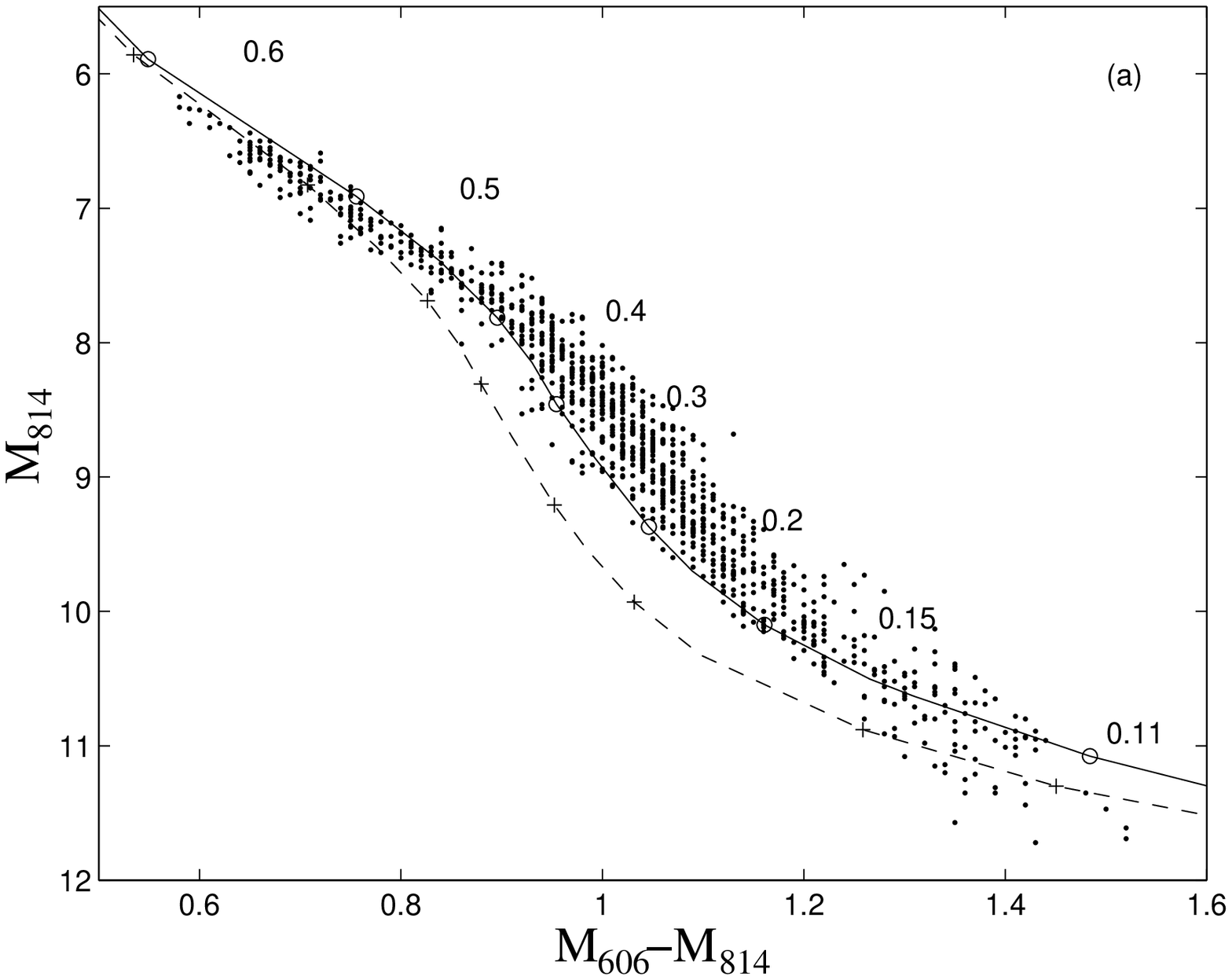} 
Fig. 5a
\end{figure*}
\begin{figure*}
\epsfxsize=160mm
\epsfysize=100mm
\epsfbox{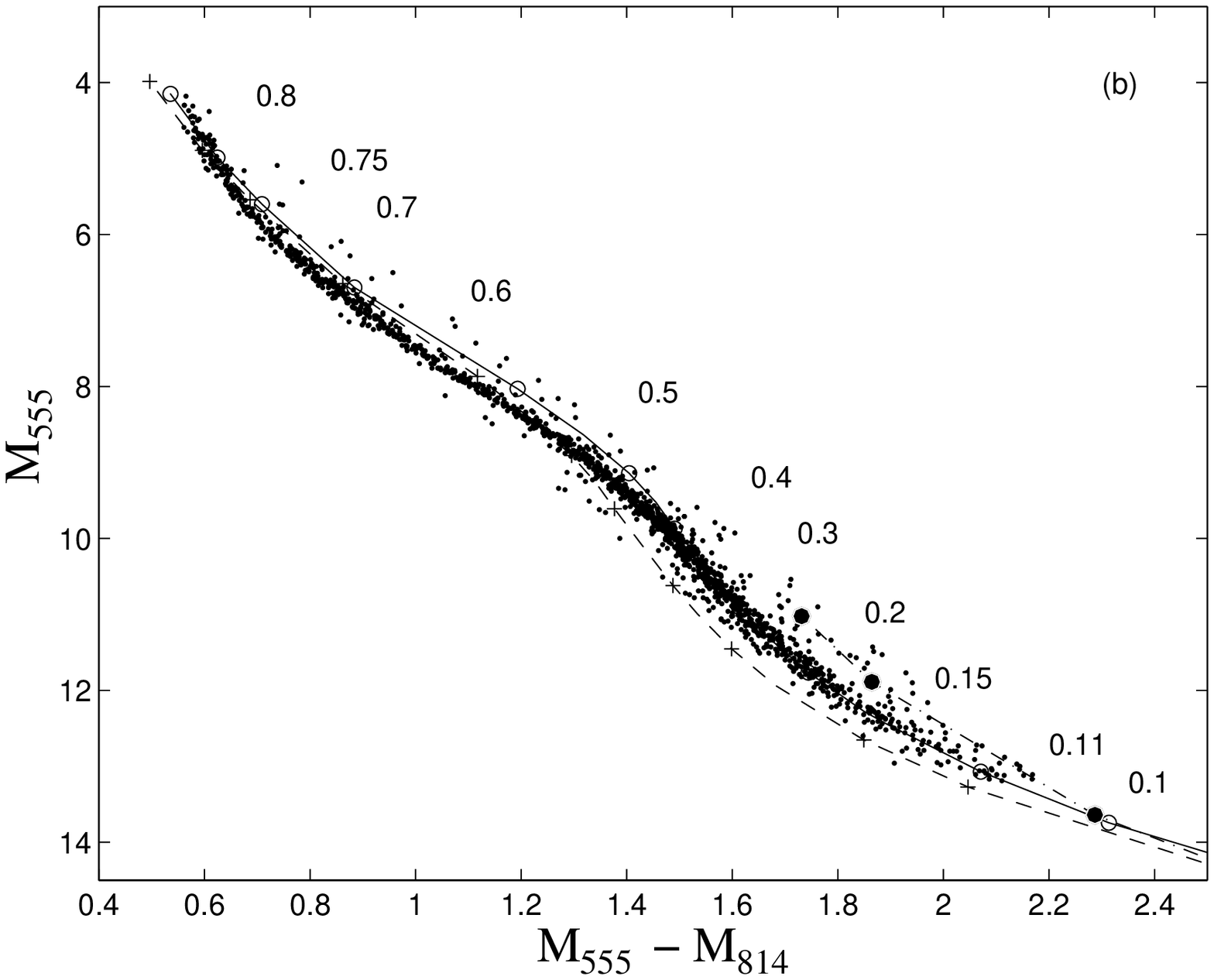}
Fig. 5b
\end{figure*}
\begin{figure*}
\epsfxsize=160mm
\epsfysize=100mm
\epsfbox{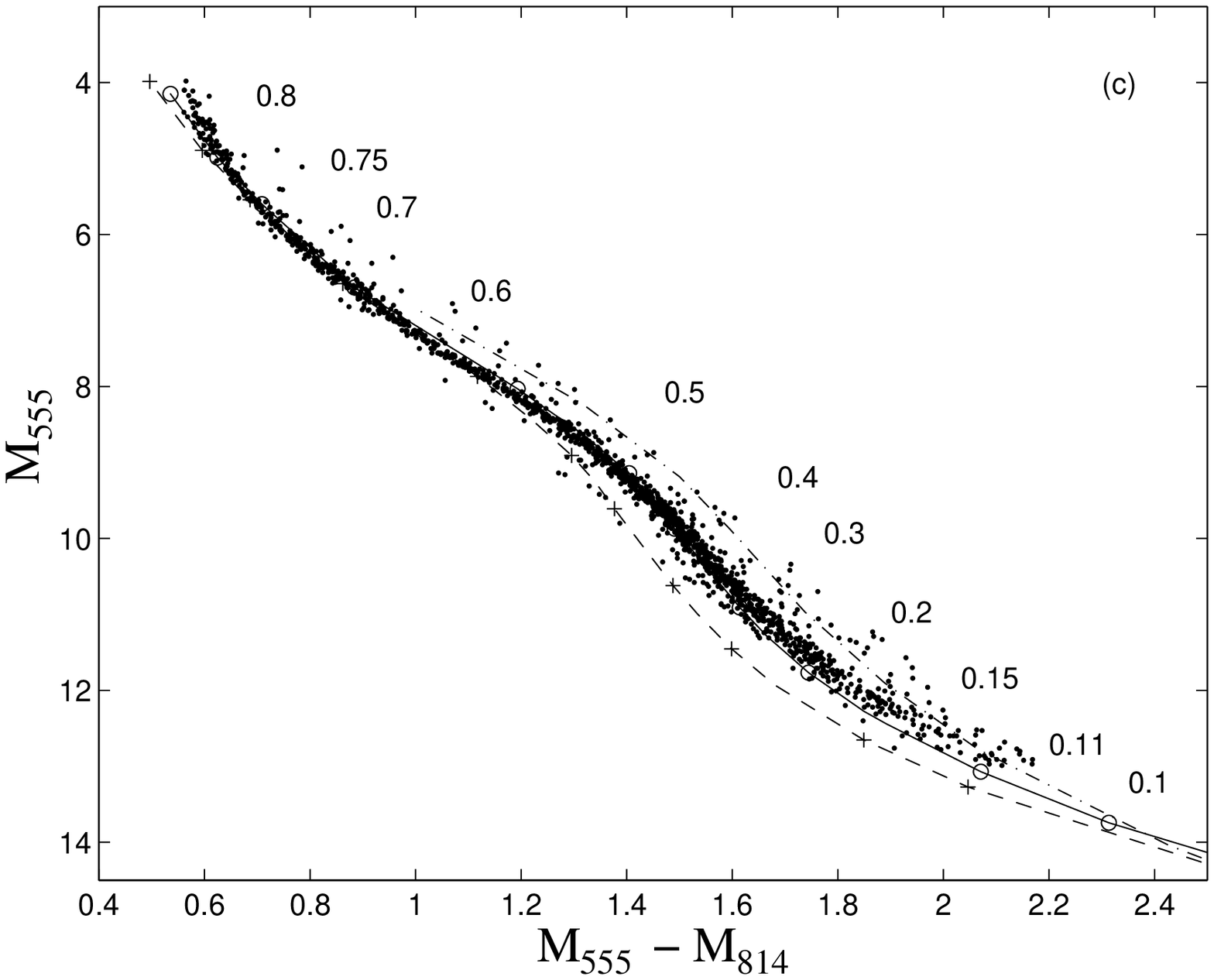}
Fig. 5c
\caption[ ]{{\bf (a)} CMD for $NGC6397$  in the $606$ and $814$ filters.
The data are from DeMarchi \& Paresce (1995). The data are dereddened
with the corrections calculated
in Table I. Solid line (o): $\mh$=-1.5; dashed line (+): $\mh$=-2.0.
The distance modulus is the one quoted by Cool et al. (1995), i.e. $(m-M)_0$=11.7.
{\bf (b)}  CMD for $NGC6397$  in the $555$ and $814$ filters.
The data are from Cool et al. (1995). The data are dereddened with the extinctions calculated
in Table I. Dot-dashed line ($\bullet$): stellar models
based on the "Base" model atmospheres (Baraffe et al. 1995) for [M/H]=-1.5; solid line (o): present stellar models based on the "NextGen" model atmospheres
for [M/H]=-1.5;
dashed line (+): $\mh$=-2.0 with the NextGen model atmospheres.
The distance modulus is the same as in Figure 5a, i.e. $(m-M)_0$=11.7.
{\bf (c)}  Same as figure 5b, with the distance modulus quoted
by DeMarchi \& Paresce (1995), i.e. $(m-M)_0$=11.9 and the same extinction as
 in Figure 5b. Solid line (o): $\mh$=-1.5; dashed line (+): $\mh$=-2.0. A comparison with the models of the Teramo group for [M/H]=-1.5 is shown (dash-dot)}
\end{figure*}

Both figures 5a and 5b are displayed with the distance modulus quoted
by Cool et al. (1996), i.e. $(m-M)_0=11.7$. The choice of the value
quoted by Paresce et al. (1995 ), $(m-M)_0=11.9$, brings the
tracks for $\mh=-1.5$ {\it exactly} on the observed sequence for the
$M_{555}$ vs $M_{555} - M_{814}$ CMD, 
 as illustrated on Figure 5c. Such an undetermination in the distance
modulus of $NGC6397$ corresponds to a difference of 200 pc, i.e. about
$10\%$. In the $M_{606}$ and $M_{814}$ filters,
the models are slightly too blue, with an offset in color $\simle 0.05$ mag,
which can well stem from the same calibration problem in $F606W$ mentioned previously for the cluster M15. The predicted sequence with the metallicity [M/H]=-1.5 remains however well within the error bars of the observed MS. 

It is interesting to analyse the agreement obtained by the Teramo
group models for this cluster.  We first note that their models, while
based on a solar-mix, correspond to $\mh=\feh$ rather than $\mh=\oh$.
As discussed in the previous section, this leads to inconsistent
comparisons, and is reflected by the fact that: i) they fit the MS
with a rather low value of the metallicity for this cluster ($-1.9 <
\mh <-1.6$), and ii) they adopt a reddening correction $E(V-I)=0.19$
which differs significantly from the value prescribed by Cool et
al. (1996) and by the present calculations (see Table I).  Using the
correct extinction would redshift significantly their models for
[M/H]=-1.5 w.r.t. the observations, as shown on Fig. 5c (dash-dotted
line).

Note that for this cluster, the observations are still
above the hydrogen-burning limit. The HBMM $\sim 0.083\,\msol$
corresponds to $\mv\sim 19.5$, $\mi \sim 13.9$ for $\mh=-1.5$, whereas
the bottom of the observed MS, $\mv\sim 13$, $\mi \sim 11$,
corresponds to $m=0.11\,\msol$. The observations of Cool et al. (1996)
extend to brighter magnitudes and reach the turn-off point, $M_V\sim
4$, $m\sim 0.8\,\pm 0.05\,\msol$.

A limited set of the first generation of the present models at
$\mh=-1.5$, based on the "Base" model atmospheres (Baraffe et
al. 1995), is also shown on Figure 5b for $m \le 0.2 \,\msol$
(full circles, dash-dot line).  These models have a slightly different
trend w.r.t. the present ones.  This stems from a general
overestimation of the molecular blanketing, the main source of
absorption in the coolest VLMS, because of the straight-mean
approximation in the "Base" models, and thus an underestimation of the
flux in the V-band ( Chabrier et al. 1996; Allard \& Hauschildt 1997).
The better agreement with the present models clearly illustrates the
recent improvements in the treatment of the molecular opacities,
especially for $TiO$ (AH97), which strongly affect the atmosphere {\it
profile}, and thus the evolution.  However, the agreement between the
previous models and observations was already quite satisfactory ($\sim
0.2$ mag) and for the first time reproduced accurately the bottom of
the observed MS for metal-poor LMS.

{$\bullet$} $\omega-Cen$ (Fig. 6) : this cluster has been observed by
Elson et al. (1995) in the F606W and F814W filters.
The observed metallicity usually quoted for this cluster is $\feh\sim -1.6$, i.e. $\mh\sim -1.3$ (cf Elson et al. 1995 and references therein), although a value
$\feh\sim -1.2$, i.e. $\mh\sim -1.0$ has been suggested recently (Norris et al.,
1996).
The
data presented in Figure 6 have been recalibrated (Elson, private
communication), since the calibrations of Holtzmann et al.  (1995)
used in Elson et al. (1995) were preliminary at this time. The zero
points for both HST filters are updated (cf. Holtzmann et al.  1995),
as well as the transformation into the Johnson-Cousins system.  Here
again, the match between theory and observation is almost perfect, as
shown in Figure 6, in {\it both} the HST instrumental and the
Jonhson-Cousins systems. Note that both systems predict the {\it same}
masses for the upper MS ($m \sim 0.75 \mso$) and the lower MS ($m \sim
0.15 \mso$), well within the error bars due to the photometric
conversion from $m_{606}$ into $V_J$ (Holtzman et al. 1995). Once again,
this assesses the accuracy of the present photometric conversions of
HST magnitudes into Johnson-Cousins magnitudes.
\begin{figure*}
\epsfxsize=180mm
\epsfysize=150mm
\epsfbox{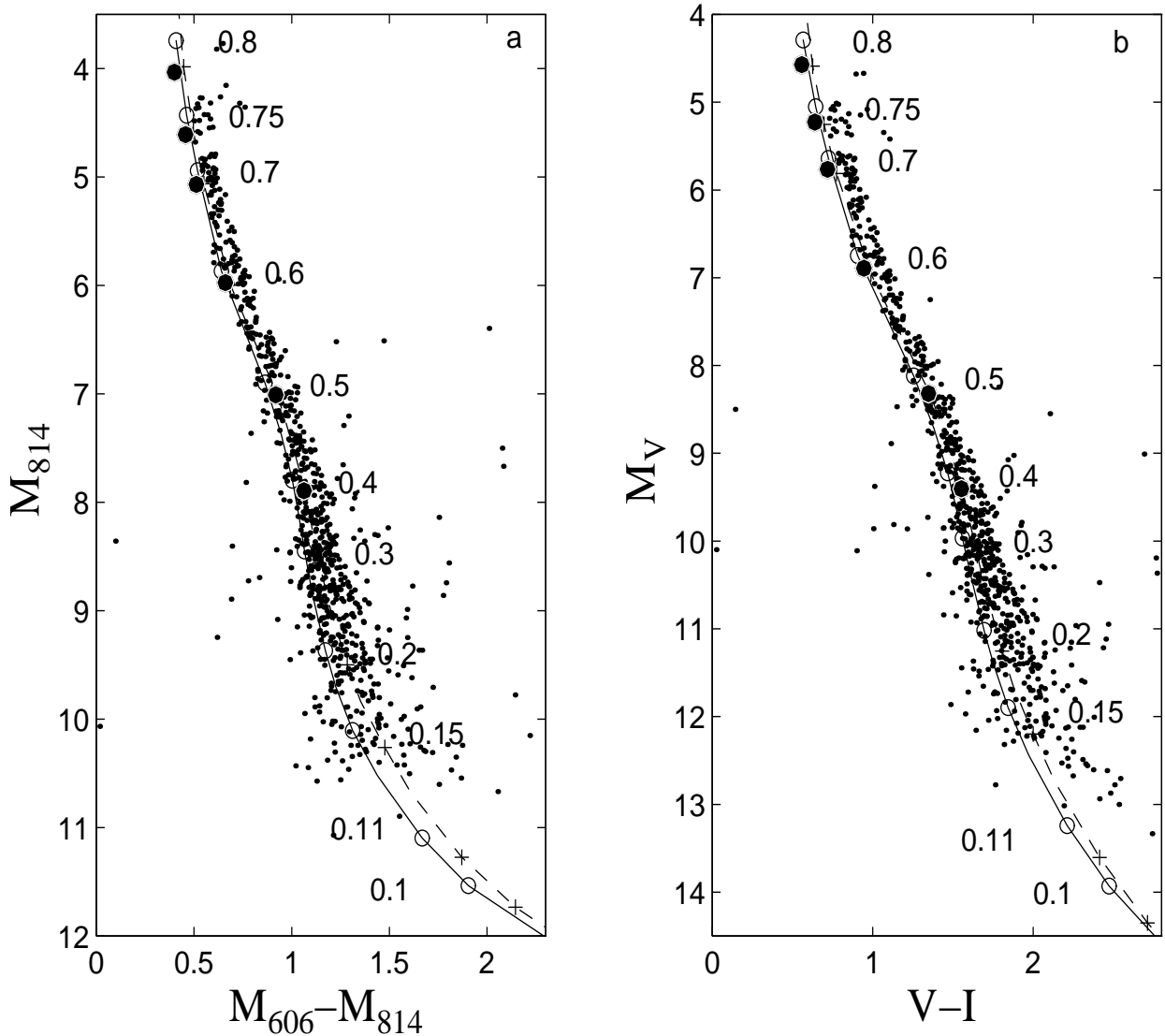} 
\caption[ ]{{\bf (a)} CMD for $\omega-Cen$. The data in $606$ and $814$ filters are from Elson et al. (1995). The models correspond to [M/H]=-1.3
(solid line, open circle) and [M/H]=-1.0 (dashed line, $+$) with $l=H_P$. Full circles
correspond to [M/H]=-1.0 and $l=2H_P$
{\bf (b)}  CMD for $\omega-Cen$ in the Johnson-Cousins system.
The data from Elson et al. (1995) have been recalibrated with updated
transformations (Elson, private communication).
Same models   as in {\bf (a)}.
}
\end{figure*}

We used the same distance modulus as Elson et al. (1995),
i.e. $(m-M)_0=13.77$.  Adopting the slightly higher value $13.92$ quoted by
Santiago et al. (1996) will shift the blue edge of the observed MS on
the theoretical isochrone with [M/H]=-1.3, yielding a less good agreement.  Our
reddening corrections (cf. Table I) are in excellent agreement with
the values quoted by Elson et al. (1995). Calculations  with a mixing length $l=2H_P$ for the [M/H]=-1 isochrone are also shown in Figure 6.
Note that the observations remain compatible with a
value $1\wig < l/H_P \wig <2$. Since the same remark applies to the lower-metallicity
clusters analysed previously, it would be hazardous to try to derive
robust conclusions about a possible dependence of the mixing length on
metallicity.
The observed width of the lower MS
does not allow a precise determination of the metallicity of the cluster from the theoretical isochrones.
The shape of the MS is well reproduced  with both
[M/H]=-1.3 and -1. This result is consistent with the spread in metallicity determined by Norris et al. (1996).

\bigskip

The properties of the models for the afore-mentioned different
metallicities, in particular the mass-color-magnitude relationships,
are displayed in Tables II-V.  The lowest mass in the table
corresponds to the hydrogen-burning minimum mass (HBMM) for each
metallicity.
For an age $t=$ 10 Gyr and
metallicities [M/H] $ \le -1.0$, the magnitude of the most massive
brown dwarf is $\sim$2 mag in $I$ and $K$, $\sim$3 mag in $V$ and
$\sim 1.5$ mag in $J$ and $H$ fainter than the one corresponding to
the HBMM. This sets the scale of the detection limit for the search
for {\it brown dwarfs} in globular clusters with future space-based
observations.

As already mentioned and clearly seen from the figures and from Tables
II - V, the more metal-depleted clusters have bluer main sequence
colors, a consequence of the increasing effective temperature at a
given mass with decreasing metallicity, since the same
optical depth corresponds to a denser layer in an increasingly
transparent atmosphere (see e.g. AH95; Chabrier \& Baraffe 1997).
Metallicity effects are most apparent in the {\it intermediate} MS,
i.e. $m \simle 0.5\,\msol$ and $\te \simle 4500$ K, where the peak of
the spectral distribution falls near the V bandpass. This stems
essentially from the increasing TiO-opacity, which absorbs mainly in
the optical
and reddens the $(V-I)$ color.  Thus, for a {\it given $\te$}, a
metal poor object will appear bluer than the more metal-rich
counterpart.  This effect is strengthened {\it at fixed mass} by the
fact that the lower the metallicity, the hotter (bluer) the star.

\subsection{Color-magnitude diagram in the near infrared}

The observations of GCs in near IR colors will soon be possible with
the next generation of HST observations, i.e. the NICMOS camera, and
in a more remote future, with the european Very-Large-Telescope (VLT).
The NICMOS filters include Wide (W), Medium (M) and Narrow (N)
bandpasses from 1.1 to 2.4 $\mu$m.  In figures 7a-b we show our
models at t=10 Gyrs for [M/H]=-2, -1.5 and -1.0 in the NICMOS wide
filters F110W, F160W and F187W. For comparison, the same isochrones
are displayed in the J ($\sim 110W$) and H ($\sim 160W$)
magnitudes (cf. Fig. 7a, dotted curves), defined in the CIT system
(Leggett 1992).  The masses listed in Table II-V are indicated by the
signs ($+$ and circles) on the curves (except 0.13 $\msol$ excluded for sake of clarity).  We note the ongoing
competition, in these infrared colors, between the reddening
due to the decreasing temperature and increasing metallic molecular
absorption in the optical and the increasing
collision-induced absorption of H$_2$ in the infrared (cf. Saumon et al.,
1994; Allard \& Hauschildt, 1995) which shifts back the flux to
shorter wavelengths. This leads to quasi-constant color
sequences from $\sim 0.5$ to $\sim 0.1$ $\mso$, corresponding to $\te
\sim 4500$ K, for which H$_2$ becomes stable in the atmosphere, to
$\sim 3500 $ K, as predicted also for zero-metallicity (Saumon et al.,
1994). Below this limit, molecular hydrogen becomes dominant,
the density keeps increasing (
cf. Chabrier \& Baraffe, 1997) and H$_2$ CIA-absorption becomes
the dominant effect ($\kappa_{CIA}\approx \kappa_{H_2-H_2}\,
\rho_{H_2}^2$ in first order, see Guillot et al., 1994).  This
causes the blue loop at the very bottom of the MS in IR colors, as
seen in Figures 7, whereas the optical colors redden
monotonically with decreasing mass, as shown in the tables. The blue
loop becomes more dramatic with decreasing metallicities, since the
lower the metallicity the denser the atmosphere.  This general trend
is very similar in all NICMOS filters covering the above-mentioned
wavelength range and is clearly a photometric signature of the stellar
to sub-stellar transition, whose physical source is the large
increase of the density in this region and the ongoing H$_2$
molecular recombination and collision-induced absorption. The limit magnitudes required
to reach the very bottom of the MS are $\sim 14$ for [M/H]=-2, $\sim
12.5$ for [M/H]=-1.5 and $\sim 12$ for [M/H]=-1, and are essentially
the same for all NICMOS filters from F110 to F240. The more massive
brown dwarfs will be about 2 magnitudes fainter.
\begin{figure*}
\epsfxsize=180mm
\epsfysize=140mm
\epsfbox{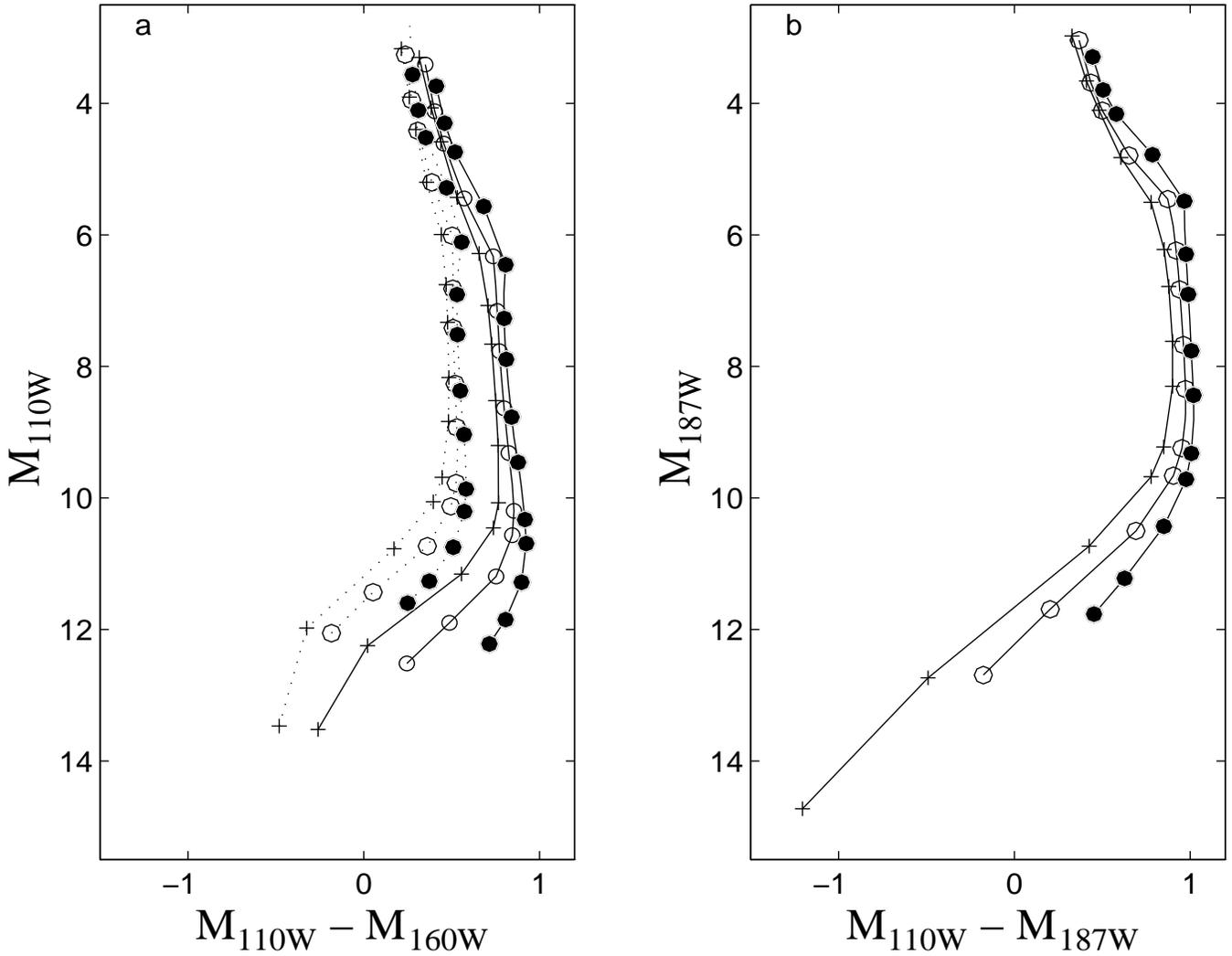} 
\caption[ ]{{\bf (a)}  Isochrones at t=10 Gyrs in the NICMOS filters
(solid curves) F110W and F160W
for three different metallicity [M/H]=-2, -1.5 and -1 (from left to right).
The dotted curves correspond to the same isochrones in the CIT
system $M_J$-(J-H). The signs on the curves correspond to the masses (except
0.13 $\msol$) 
tabulated in Tables II to V. {\bf (b)} Same as in Fig. 7a in the NICMOS filters F110W and
F187W.
}
\end{figure*}

\subsection{Comparison with the halo subdwarf sequence}

Figure 8 displays different observations of halo field stars in the
standard Johnson-Cousins system. The filled circles are the {\it
subdwarf} sequence of Monet et al.  (1992), the crosses the more
complete subdwarf sequence of Dahn et al. (1995), and the triangles
correspond to a sub-sample of Leggett's (1992) halo stars. The halo
classification was determined photometrically (Leggett 1992) and
kinematically : the stars in the three afore-mentioned samples have
tangential velocities $|V_{tan}|>220$ km.s$^{-1}$ (Monet et al.),
$>$160 km.s$^{-1}$ (Dahn et al.)  and $>$180 km.s$^{-1}$ (present
sub-sample of Leggett). All these observations appear to be fairly
consistent, the Monet et al. sample representing the most extreme
halo fraction of the Dahn et al. sample, the Leggett's sample containing
only a few genuine halo stars. We stress that the linear fit proposed
by Leggett (1992) is a rather poor representation of the distribution
of the true halo objects and is strongly misleading. We also emphasize
that linear fits are {\it not} correct to fit LMS sequences in the HR
diagram, since they do not reproduce the wavy behaviour of the
sequences, wich reflects intrinsic {\it physical} properties of these
stars, as discussed in \S 3.1.  This non linearity has already been
stressed for the mass-luminosity (D'Antona \& Mazzitelli, 1994;
Chabrier et al., 1996) and the mass-spectral class relations (Baraffe
\& Chabrier, 1996).  We first note the important spread in color, over
1 mag, which reflects the large spread in metallicity in the sample.

\begin{figure*}
\epsfxsize=160mm
\epsfysize=120mm
\epsfbox{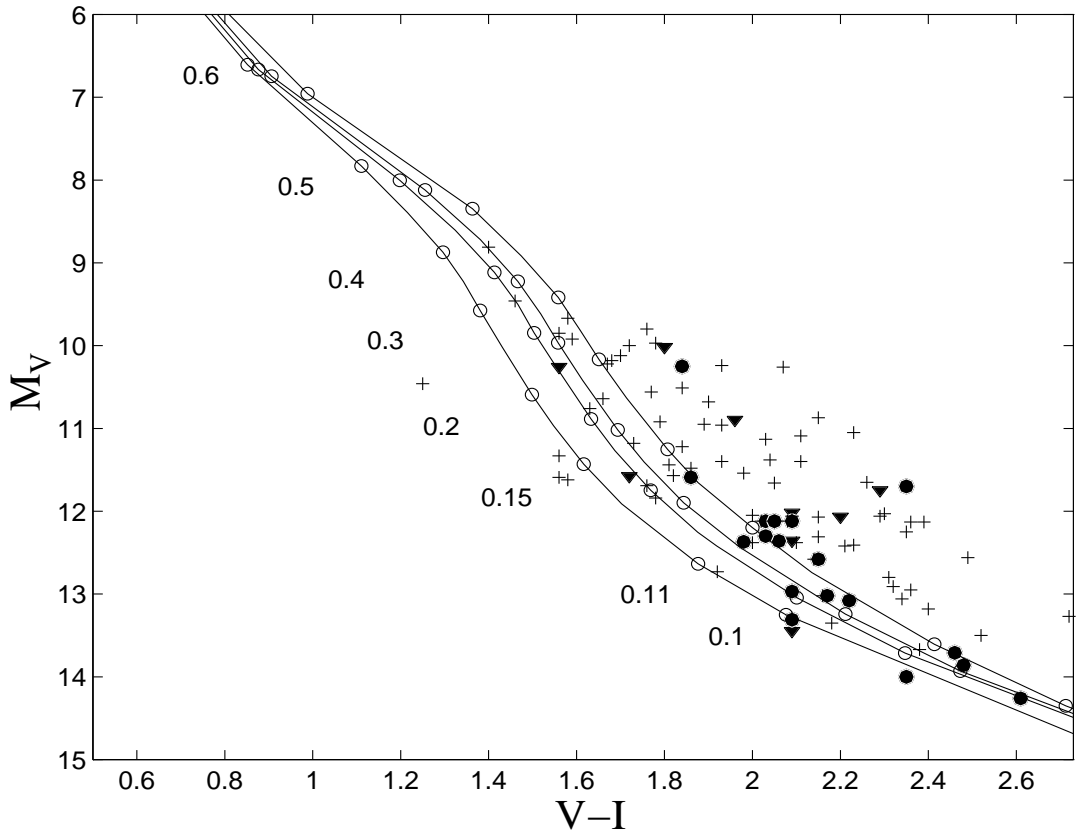} 
\caption[ ]{Subdwarf halo field stars
from the data of Monet et al. (1992) (full circles), Dahn et
al. (1995) ($+$) and Leggett (1992) (triangles), in the standard
Jonhson-Cousins system.  All the stars shown here from the Leggett's
sample have tangential velocities $|V_{tan}|\ge 180$ km.s$^{-1}$.
Solid lines: present models for $\mh$=-2,-1.5, -1.3 and -1 (from left
to right) for t=10 Gyr. The empty circles correspond to the masses
indicated on the figure.}
\end{figure*}

The solid lines indicate the present LMS sequences
for $\mh=-2.0,\,-1.5,\, -1.3$ and -1.0 from left to right.
The Monet et al. sample is
consistent with an {\it average} metallicity $\mh\sim -1.3$ to $-1.5$. It is more
hazardous to try to infer an average value for Leggett's sample, given the
limited number of objects and the large dispersion. The Dahn at al. sample
clearly includes the two previous ones, with objects ranging from $\mh\sim -0.5$
to $\mh=-2.0$. This reflects 
the difficulty to determine the precise origin of an object
from its kinematic properties only. 

As seen on the figure, these field stars are fairly consistent with the theoretical sequences determined for the clusters, for similar metallicities. This is particularly obvious for the sequences of
$NGC6397$ ([M/H]=-1.5) and $\omega \,Cen$ ([M/H]=-1.3), which match perfectly
the Monet et al. average sequence. This strongly suggests,
contrarily to what has been suggested by Santiago et al. (1996),
that there is no significant difference in the structure and the
evolution of globular star clusters and field halo stars.
The discrepancy between the CMD of Leggett (1992)
and those from the HST led Santiago et al. (1996) to invoke the possibility
of a calibration problem of the HST data. However, the agreement that we find
between our theoretical models and the globular cluster CMDs in both the HST
data {\it and} the Johnson-Cousins system (e.g for $M15$ and $wCen$), excludes this hypothesis.
It rather stems form the large metallicity dispersion in Leggett's sample and from the
misleading fit of this sample.

\section{Conclusion}

We have presented extensive calculations of VLMS evolution
in the range $-2.0\le \mh \le -1.0$, characteristic metallicities for old globular star clusters and halo field stars.  These
low-metallicities minimize possible shortcomings of the model
atmospheres pertaining to incomplete or inaccurate metallic molecular line
lists and grain formation, and provide a stepping stone towards the derivation of more
accurate low-mass star models for solar metallicity.  The models are
examined against available deep photometry color-magnitude diagrams
obtained with the Hubble Space Telescope for three globular clusters.  The
HST CMDs for the clusters span a large range in metallicity, thus
providing very stringent tests for the models.  Since the parameters
characteristic of these clusters, extinction, distance modulus and
metallicity are fairly well defined, there is no free parameter left
to bring models into agreement with observations.  Therefore
comparison between theory and observation reflects directly the accuracy of the theory.
We stress the importance of eqn.(1) when comparing GC observed and
theoretical CMDs. A first generation of the present models (Baraffe
et al. 1995) has been used incorrectly by comparing observations at
$\feh$ with a model at the same value of $\mh$.

The main conclusions of these calculations can be summarized as follows :


{$\bullet$}
We first note the overall remarkable agreement between the present
models and the
observations, within less than 0.1 mag, over the {\it whole} metallicity
range and the entire main sequence from the turn-off to the bottom. The characteristic changes in the slopes of the cluster MS's are
reproduced accurately, and assess the validity of the
physics involved in the models.
This yields an accurate calibration of the observations, i.e. reliable
mass-magnitude-effective temperature-age relationships.

We also provide reddening corrections based on accurate
LMS synthetic spectra.

{$\bullet$}
Variations of the mixing length in the stellar {\it interior}
affect essentially the upper main sequence near the turn-off, i.e. only
stars
massive enough to develop a large radiative core ($m\wig > 0.7 \,\msol$).
Variations of the mixing length in the {\it atmosphere} is found to
be inconsequential on evolutionary models.
There is no clear hint for a dependence
of the mixing length on the metallicity and our results remain in agreement
with all observed sequences, within the error bars, for $1\le l_{mix}/H_P \le 2$.

{$\bullet$}
The Ryan \& Norris (1991) prescription to convert solar-mix
abundances into oxygen-enriched mixtures characteristic of old stellar
populations is accurate. Using the correct $\alpha$-enhanced mixture
(both in the atmosphere and in the interior opacities) leads to results
identical to those obtained with the afore-mentioned scaling.
Not taking this
enrichment into account leads to inconsistent comparisons.


{$\bullet$}
We derive theoretical sequences in the filters of the
NICMOS camera, down to the hydrogen burning limit. This will allow
a straightforward analysis of the future HST observations, and provides
a stringent test for the accuracy of the present models near the
brown dwarf limit. This corresponds to $M_J\approx M_H\approx M_K\approx
13-14$ for the lowest metallicity
 examined presently, i.e.
$\mh=-1.0$.

We also predict a photometric signature of the transition from stellar
to substellar objects in the infrared, in terms of a severe blue loop
near the very bottom of the MS, whereas optical colors keep reddening
almost linearly. This photometric signature reflects the overwhelming
absorption of molecular hydrogen in the infrared due to many-body collisions,
and stems from the increase of the molecular hydrogen fraction and of the
density (contraction) near the
stellar to substellar transition.

{$\bullet$}
The models allow a good determination of the metallicty of the
observed halo {\it field} stars.
A striking result is the large metallicity dispersion of these objects,
from $\mh$ = -2 to near solar, although they all have halo kinematic properties.
 The most extreme
halo stars, represented by the Monet et al. (1992) sample, have a metallicity
ranging from $\mh\sim -1.0$ to -2.0, with an average value $\mh\sim -1.3$.
We find no evidence for differences in the sequences of
halo field subdwarfs and globular star clusters.
Both are reproduced with the same isochrones for
similar {\it mean} metallicity, $\mh \sim -1.3$ to $-1.5$, i.e. $\feh\sim
-1.6$ to $-1.8$.


We can now affirm that the theory of low-mass stars, at least for
metal-depleted abundances, has reached a very good level of accuracy and
can be used with confidence to analyse the observations and make reliable
predictions.  While we cannot honestly exclude the possibility of 
a discrepancy in the $M_{606} - M_{814}$ color predicted by the models
(by $\sim 0.05 $ mag), it remains within the error bars due to the
observations themselves and to the undeterminations of either the
extinction or the distance modulus. It may also reflect present uncertainties in
the calibration of the $F606$ filter.

While it is unfortunate that the hydrogen burning limit remains
unreached by the present HST observations, for it is masked by
foreground field stars, the next cycles of HST observations will be
able to resolve it in the near future.  Indeed, known proper motion
of the GCs will cause a substantial displacement of cluster stars
between 1994 and 1997 which should allow a separation of cluster
from field stars (King 1995). Also, as mentioned above, observations
in the IR should lead to the observation of the very bottom of - and
possibly below - the MS. The present models provide the
limit magnitudes to be reached to enter the
brown dwarf regime.

The present calculations represent an important improvement in the
description of the mechanical and thermal properties of low-mass
stars, and of their photometric signature.  This provides solid
grounds to extend these calculations into the more complicated domain
of solar-like metallicities, as will be examined in a forthcoming
paper  (Allard et al., 1997b).  The assessed accuracy of the present
models provide reliable mass-luminosity relationships for metal-poor
stellar populations in general and for globular clusters in
particular.  This allows for the first time the derivation of {\it
reliable mass-functions} for these objects down to the brown dwarf
limit (Chabrier and M\'era, 1997).

\medskip
Tables II-V are available by anonymous ftp: 
\par
\hskip 1cm ftp ftp.ens-lyon.fr \par
\hskip 1cm username: anonymous \par
\hskip 1cm ftp $>$ cd /pub/users/CRAL/ibaraffe \par
\hskip 1cm ftp $>$ get BCAH97\_models \par
\hskip 1cm ftp $>$ quit
\bigskip
\bigskip

\begin{acknowledgements} The authors are very grateful to A. Cool,
I. King, F. Paresce, G. DeMarchi, R. Elson and G. Gilmore, for kindly
providing their data under electronic forms.  We are particularly
indebted to A. Cool, G. DeMarchi, F. Paresce, and J. Holtzman for
sending us the WFPC2 555, 814 and 606 transmission
filters and additional transformations, and to F. Rogers
for computing opacities with $\alpha$-enriched abundances upon request.
We also thank C. Dahn for sending his tabulated data.

This work is funded by grants from NSF AST-9217946 to Indiana
University and NASA LTSA NAG5-3435 and NASA EPSCoR NAG-3435 to Wichita State University.
PHH was partially supported by
NASA LTSA and ATP grants to the University of Georgia in Athens.
The calculations presented in this paper were performed at the Cornell
Theory Center (CTC), the San Diego Supercomputer Center (SDSC), and on the
CRAY C90 of the Centre d'Etudes Nucl\'eaire
de Grenoble.
\end{acknowledgements}

\vfill
\eject

\begin{table}
\caption{Characteristics of the globular clusters. $[Fe/H]$ is the observed metallicity, $[M/H]$ is the metallicity used in the models, which takes into
account the $\alpha$-enrichment following the prescription of
Ryan \& Norris (see \S 2). The fourth and fifth
columns denote the bare distance modulus and reddening repectively, as quoted by the refered observers.
The last three columns give the extinction in the three respective filters
$F555 \sim V_J$, $F606$ and $F814 \sim I_C$,
where $J$ and $C$ denote the standard Johnson-Cousins system. The upper
rows give the value chosen by the refered observers whereas the lower
rows denote the value used in the present calculations, deduced from the
Allard \& Hauschildt (1997) synthetic spectra and the Cardelli et al. (1989)
extinction law.
In all cases the interstellar extinction corresponds to $R_V=3.12$.}
\begin{tabular}{llllllll}\hline
\hline\noalign{\smallskip}
$cluster$ & $[Fe/H]$ & $[M/H]$
& $(m-M)_0$ & E(B-V) &$A_{555}$ & $A_{606}$ & $A_{814}$ \\
\noalign{\smallskip}
\hline\noalign{\smallskip}
 M15$^{1,2}$      & -2.3    & -2.0 & 15.10 & 0.11 & 0.34 & 0.274 & 0.144 \\
                  &         &      & 15.10 & 0.11 & 0.33 & 0.30 & 0.20 \\

 NGC6397$^{3}$    & -1.9    & -1.5 & 11.70 & 0.18 &0.57 &      & 0.34\\
 NGC6397$^{4}$    & -1.9    & -1.5 & 11.90 &  0.18 & &      &  0.30\\
                  &         &      & 11.7 (11.9)& 0.18&0.57 & 0.51 & 0.35 \\

 $\omega$ Cen$^5$ & -1.6    & -1.3/-1.0 & 13.77 & 0.116&0.36 & 0.32 & 0.22\\
                  &         &     & 13.77 & 0.116&0.36 & 0.33 & 0.22\\
%
\hline
\end{tabular}
\bigskip


$^1$ Piotto et al. 1995

$^2$ De Marchi \& Paresce 1995

$^3$ Cool et al., 1996

$^4$ Paresce et al. 1995

$^5$ Elson et al. 1995

\end{table}

\begin{table*}
\caption{Physical properties and absolute magnitudes of low-mass stars for
$[M/H] = -2.0$  and $\tau$ = 10 Gyrs. The lowest mass corresponds to the hydrogen-burning limit. The mass $m$ is in $M_\odot$, $T_{eff}$ in K and the luminosity L in $L_\odot$. The VRI magnitudes are in the Johnson-Cousins system (Bessell 1990) and the JHK magnitudes in the CIT system (Leggett 1992).
Note that the bolometric magnitude corresponds to $M_{bol}(\odot)$ = 4.64.}
\begin{tabular}{lcccccccccc}
\hline\noalign{\smallskip}
$m$  &$T_{eff}$&log L & log g &$M_{bol}$ &$M_V$ &$M_R$ &$M_I$ &$M_J$ &
$M_H$ & $M_K$ \\
\noalign{\smallskip}
\hline\noalign{\smallskip}
 0.083 & 1779. & -4.274& 5.58 & 15.33 & 20.92 & 16.92 & 14.41 & 13.47 & 13.95 & 14.58\\
 0.085 & 2342. & -3.741& 5.53 & 13.99 & 17.67 & 15.17 & 13.11 & 11.98 & 12.30 & 12.61\\
 0.090 & 3005. & -3.202& 5.45 & 12.65 & 14.70 & 13.40 & 11.96 & 10.77 & 10.60 & 10.68\\
 0.100 & 3395. & -2.860& 5.37 & 11.79 & 13.25 & 12.22 & 11.17 & 10.06 &  9.66 &  9.58\\
 0.110 & 3565. & -2.686& 5.32 & 11.36 & 12.63 & 11.66 & 10.76 &  9.69 &  9.24 &  9.12\\
 0.130 & 3749. & -2.463& 5.26 & 10.80 & 11.91 & 10.99 & 10.21 &  9.20 &  8.72 &  8.58\\
 0.150 & 3852. & -2.306& 5.21 & 10.40 & 11.43 & 10.55 &  9.81 &  8.84 &  8.36 &  8.20\\
 0.200 & 4004. & -2.020& 5.12 &  9.69 & 10.59 &  9.78 &  9.09 &  8.17 &  7.69 &  7.52\\
 0.300 & 4172. & -1.665& 5.01 &  8.80 &  9.58 &  8.83 &  8.19 &  7.33 &  6.85 &  6.69\\
 0.400 & 4304. & -1.421& 4.94 &  8.19 &  8.87 &  8.18 &  7.58 &  6.75 &  6.29 &  6.13\\
 0.500 & 4623. & -1.083& 4.83 &  7.35 &  7.83 &  7.25 &  6.72 &  5.99 &  5.55 &  5.43\\
 0.600 & 5292. & -0.676& 4.73 &  6.33 &  6.61 &  6.17 &  5.76 &  5.20 &  4.84 &  4.77\\
 0.700 & 5914. & -0.265& 4.58 &  5.30 &  5.51 &  5.17 &  4.83 &  4.40 &  4.10 &  4.07\\
 0.750 & 6282. & -0.015& 4.47 &  4.68 &  4.87 &  4.57 &  4.28 &  3.90 &  3.64 &  3.63\\
 0.800 & 6688. &  0.334& 4.25 &  3.81 &  3.96 &  3.72 &  3.47 &  3.17 &  2.95 &  2.96\\
\hline
\end{tabular}
\end{table*}

\begin{table*}
\caption{Same as in Table 2 for $[M/H] = -1.5$  }
\begin{tabular}{lcccccccccc}
\hline\noalign{\smallskip}
$m$  &$T_{eff}$&log L & log g &$M_{bol}$ &$M_V$ &$M_R$ &$M_I$ &$M_J$ &
$M_H$ & $M_K$ \\
\noalign{\smallskip}
\hline\noalign{\smallskip}
 0.083 & 2194. & -3.844& 5.51 & 14.25 & 19.42 & 16.32 & 13.87 & 12.06 & 12.24 & 12.42\\
 0.085 & 2519. & -3.559& 5.48 & 13.54 & 17.50 & 15.20 & 13.05 & 11.43 & 11.38 & 11.53\\
 0.090 & 2938. & -3.207& 5.42 & 12.66 & 15.21 & 13.78 & 12.13 & 10.73 & 10.37 & 10.37\\
 0.100 & 3255. & -2.910& 5.34 & 11.92 & 13.71 & 12.61 & 11.37 & 10.13 &  9.63 &  9.52\\
 0.110 & 3405. & -2.745& 5.30 & 11.50 & 13.04 & 12.02 & 10.94 &  9.77 &  9.25 &  9.10\\
 0.130 & 3586. & -2.523& 5.24 & 10.95 & 12.25 & 11.29 & 10.38 &  9.28 &  8.75 &  8.57\\
 0.150 & 3691. & -2.366& 5.20 & 10.56 & 11.75 & 10.81 &  9.98 &  8.93 &  8.40 &  8.21\\
 0.200 & 3845. & -2.082& 5.11 &  9.84 & 10.89 & 10.00 &  9.25 &  8.26 &  7.74 &  7.55\\
 0.300 & 4008. & -1.724& 5.00 &  8.95 &  9.85 &  9.03 &  8.34 &  7.42 &  6.91 &  6.72\\
 0.400 & 4134. & -1.471& 4.92 &  8.32 &  9.11 &  8.35 &  7.70 &  6.82 &  6.31 &  6.14\\
 0.500 & 4462. & -1.116& 4.80 &  7.43 &  8.00 &  7.36 &  6.80 &  6.01 &  5.51 &  5.39\\
 0.600 & 5167. & -0.702& 4.72 &  6.40 &  6.66 &  6.21 &  5.79 &  5.20 &  4.81 &  4.75\\
 0.700 & 5792. & -0.296& 4.58 &  5.38 &  5.57 &  5.21 &  4.87 &  4.42 &  4.11 &  4.08\\
 0.750 & 6132. & -0.059& 4.47 &  4.79 &  4.96 &  4.65 &  4.34 &  3.95 &  3.67 &  3.66\\
 0.800 & 6483. &  0.266& 4.27 &  3.98 &  4.12 &  3.86 &  3.59 &  3.26 &  3.02 &  3.02\\
\hline
\end{tabular}
\end{table*}

\begin{table*}
\caption{Same as in Table 2 for $[M/H] = -1.3$  }
\begin{tabular}{lcccccccccc}
\hline\noalign{\smallskip}
$m$  &$T_{eff}$&log L & log g &$M_{bol}$ &$M_V$ &$M_R$ &$M_I$ &$M_J$ &
$M_H$ & $M_K$ \\
\noalign{\smallskip}
\hline\noalign{\smallskip}
 0.083 & 2283. & -3.751& 5.49 & 14.02 & 19.24 & 16.30 & 13.83 & 11.81 & 11.82 & 11.95\\
 0.085 & 2542. & -3.524& 5.46 & 13.45 & 17.61 & 15.32 & 13.12 & 11.34 & 11.15 & 11.25\\
 0.090 & 2903. & -3.214& 5.40 & 12.68 & 15.47 & 13.98 & 12.23 & 10.74 & 10.31 & 10.27\\
 0.100 & 3200. & -2.931& 5.34 & 11.97 & 13.93 & 12.79 & 11.46 & 10.16 &  9.63 &  9.49\\
 0.110 & 3343. & -2.769& 5.29 & 11.56 & 13.24 & 12.20 & 11.03 &  9.81 &  9.26 &  9.09\\
 0.130 & 3512. & -2.552& 5.23 & 11.02 & 12.44 & 11.46 & 10.47 &  9.33 &  8.77 &  8.58\\
 0.150 & 3624. & -2.392& 5.19 & 10.62 & 11.90 & 10.95 & 10.05 &  8.97 &  8.42 &  8.22\\
 0.200 & 3780. & -2.107& 5.10 &  9.91 & 11.02 & 10.11 &  9.32 &  8.30 &  7.77 &  7.57\\
 0.300 & 3945. & -1.747& 4.99 &  9.01 &  9.97 &  9.12 &  8.41 &  7.46 &  6.94 &  6.74\\
 0.400 & 4067. & -1.492& 4.92 &  8.37 &  9.22 &  8.42 &  7.76 &  6.85 &  6.33 &  6.16\\
 0.500 & 4377. & -1.139& 4.79 &  7.49 &  8.12 &  7.44 &  6.86 &  6.04 &  5.51 &  5.39\\
 0.600 & 5068. & -0.727& 4.71 &  6.46 &  6.75 &  6.27 &  5.84 &  5.22 &  4.81 &  4.74\\
 0.700 & 5686. & -0.326& 4.58 &  5.46 &  5.64 &  5.27 &  4.92 &  4.44 &  4.12 &  4.09\\
 0.750 & 6006. & -0.099& 4.47 &  4.89 &  5.05 &  4.72 &  4.41 &  3.99 &  3.71 &  3.69\\
 0.800 & 6315. &  0.199& 4.29 &  4.14 &  4.29 &  4.00 &  3.72 &  3.36 &  3.11 &  3.10\\
\hline
\end{tabular}
\end{table*}


\begin{table*}
\caption{Same as in Table 2 for $[M/H] = -1.0$  }
\begin{tabular}{lcccccccccc}
\hline\noalign{\smallskip}
$m$  &$T_{eff}$&log L & log g &$M_{bol}$ &$M_V$ &$M_R$ &$M_I$ &$M_J$ &
$M_H$ & $M_K$ \\
\noalign{\smallskip}
\hline\noalign{\smallskip}
 0.083 & 2359. & -3.660& 5.45 & 13.79 & 19.10 & 16.42 & 13.93 & 11.60 & 11.35 & 11.42\\
 0.085 & 2550. & -3.492& 5.43 & 13.37 & 17.81 & 15.58 & 13.31 & 11.26 & 10.89 & 10.91\\
 0.090 & 2840. & -3.231& 5.38 & 12.72 & 15.94 & 14.35 & 12.44 & 10.75 & 10.24 & 10.15\\
 0.100 & 3107. & -2.967& 5.32 & 12.06 & 14.35 & 13.15 & 11.64 & 10.21 &  9.63 &  9.46\\
 0.110 & 3245. & -2.808& 5.28 & 11.66 & 13.61 & 12.52 & 11.19 &  9.87 &  9.28 &  9.09\\
 0.130 & 3406. & -2.593& 5.22 & 11.12 & 12.74 & 11.74 & 10.61 &  9.39 &  8.81 &  8.60\\
 0.150 & 3505. & -2.439& 5.18 & 10.74 & 12.20 & 11.23 & 10.20 &  9.04 &  8.46 &  8.25\\
 0.200 & 3674. & -2.150& 5.10 & 10.02 & 11.25 & 10.33 &  9.45 &  8.37 &  7.82 &  7.60\\
 0.300 & 3841. & -1.787& 4.99 &  9.11 & 10.17 &  9.28 &  8.51 &  7.52 &  6.98 &  6.78\\
 0.400 & 3959. & -1.530& 4.91 &  8.46 &  9.42 &  8.57 &  7.86 &  6.91 &  6.38 &  6.18\\
 0.500 & 4235. & -1.185& 4.78 &  7.60 &  8.35 &  7.60 &  6.98 &  6.11 &  5.55 &  5.42\\
 0.600 & 4867. & -0.782& 4.69 &  6.60 &  6.96 &  6.42 &  5.97 &  5.28 &  4.81 &  4.74\\
 0.700 & 5478. & -0.392& 4.58 &  5.62 &  5.81 &  5.41 &  5.04 &  4.52 &  4.17 &  4.13\\
 0.750 & 5776. & -0.181& 4.49 &  5.09 &  5.25 &  4.89 &  4.56 &  4.10 &  3.79 &  3.77\\
 0.800 & 6055. &  0.077& 4.34 &  4.45 &  4.59 &  4.27 &  3.96 &  3.56 &  3.28 &  3.27\\
\hline
\end{tabular}
\end{table*}

\end{document}